\documentclass[amsfonts,aps,prb]{revtex4}

\def\onehalf{\textstyle{\frac{1}{2}}}
\def\D{{\mathcal D}{}}
\def\Gammabol{{\stackrel{\circ}{\Gamma}}{}}

\def\Abol{{\stackrel{~\circ}{A}}{}}
\def\Rbol{{\stackrel{\circ}{R}}{}}
\def\Lbol{{\stackrel{\circ}{\mathcal L}}{}}

\def\Dbol{{\stackrel{\circ}{\mathcal D}}{}}
\def\nabol{{\stackrel{\circ}{\nabla}}{}}

\def\Gammaw{{\stackrel{\bullet}{\Gamma}}{}}
\def\Omegaw{{\stackrel{\bullet}{\Omega}}{}}
\def\Aw{{\stackrel{~\bullet}{A}}{}}
\def\Vw{{\stackrel{\bullet}{V}}{}}
\def\Rw{{\stackrel{\bullet}{R}}{}}
\def\Qw{{\stackrel{\bullet}{Q}}{}}
\def\jw{{\stackrel{\bullet}{\jmath}}{}}
\def\tw{{\stackrel{\bullet}{t}}{}}

\def\Lw{{\stackrel{\bullet}{\mathcal L}}{}}
\def\Tw{{\stackrel{\bullet}{T}}{}}
\def\Kw{{\stackrel{\bullet}{K}}{}}
\def\nablaw{{\stackrel{\bullet}{\nabla}}{}}
\def\Dw{{\stackrel{\bullet}{\mathcal D}}{}}
\def\dw{{\stackrel{\bullet}{D}}{}}
\def\Sw{{\stackrel{\bullet}{S}}{}}
\def\be{\begin{equation}}
\def\ee{\end{equation}}
\def\ba{\begin{eqnarray}}
\def\ea{\end{eqnarray}}
\def\nn{\nonumber}
\def\qed{\hbox{${\vcenter{\vbox{
   \hrule height 0.4pt\hbox{\vrule width 0.4pt height 6pt
   \kern5pt\vrule width 0.4pt}\hrule height 0.4pt}}}$}}
\def\qedw{{\stackrel{\bullet}{\qed}}{}}

\begin{document}

\title{TORSION GRAVITY: A REAPPRAISAL}
\author{H. I. Arcos\footnote{Permanent address: Universidad Tecnol\'ogica de Pereira, A.A. 97,
La Julita, Pereira, Colombia}}
\author{J. G. Pereira}
\affiliation{Instituto de F\'{\i}sica Te\'orica, Universidade Estadual Paulista \\ 
Rua Pamplona 145, 01405-900 S\~ao Paulo, Brazil}

\begin{abstract}
The role played by torsion in gravitation is critically reviewed. After a description of the
problems and controversies involving the physics of torsion, a comprehensive presentation of
the teleparallel equivalent of general relativity is made. According to this theory,
curvature and torsion are alternative ways of describing the gravitational field, and
consequently related to the same degrees of freedom of gravity. However, more general gravity
theories, like for example Einstein--Cartan and gauge theories for the Poincar\'e and the
affine groups, consider curvature and torsion as representing independent degrees of freedom.
By using an {\em active} version of the strong equivalence principle, a possible solution to
this conceptual question is reviewed. This solution favors ultimately the teleparallel point
of view, and consequently the completeness of general relativity. A discussion of the
consequences for gravitation is presented.
\end{abstract}

\maketitle

\section{Introduction}

\subsection{General Concepts: Riemann versus Weitzenb\"ock}

Gravitation presents a quite peculiar property: particles with different masses and
different compositions feel it in such a way that all of them acquire the same acceleration
and, given the same initial conditions, follow the same path. Such universality of
response---usually referred to as {\it universality of free fall}---is the most fundamental
characteristic of the gravitational interaction.\cite{exp} It is a unique property, peculiar
to gravitation: no other basic interaction of Nature has it. Effects equally felt by all
bodies were known since long. They are the so called {\it inertial} effects, which show up
in non-inertial frames. Examples on Earth are the centrifugal and the Coriolis forces.

Universality of inertial effects was one of the hints used by Einstein towards general
relativity, his theory for gravitation. Another ingredient was the notion of field. The
concept allows the best approach to interactions coherent with special relativity. All known
forces are mediated by fields on spacetime. If gravitation is to be represented by a field,
it should, by the considerations above, be a universal field, equally felt by every particle.
A natural solution is to assume that gravitation changes the spacetime itself. And, of all
the fields present in a spacetime, the metric appears as the most fundamental. The simplest
way to change spacetime, then, would be to change its metric. Furthermore, the metric does
change when looked at from a non-inertial frame, where the (also universal) inertial effects
are present. According to this approach, therefore, the gravitational field should be
represented by the spacetime metric. In the absence of gravitation, the spacetime metric
should reduce to the flat Minkowski metric.

A crucial point of this description of the gravitational interaction, which is a description
fundamentally based on the universality of free fall, is that it makes no use of the concept
of {\em force}. In fact, according to it, instead of acting through a force, the presence of
gravitation is represented by a deformation of the spacetime structure. More precisely, the
presence of a gravitational field is supposed to produce a {\em curvature} in spacetime, the
gravitational interaction being achieved by letting (spinless) particles to follow the
geodesics of the underlying spacetime. Notice that no other kind of spacetime deformation is
supposed to exist. Torsion, for example, which would be another natural spacetime
deformation, is assumed to vanish from the very beginning. This is the approach of general
relativity, in which geometry replaces the concept of gravitational force, and the
trajectories are determined, not by force equations, but by geodesics.\cite{mtw} The
spacetime underlying this theory is a pseudo Riemannian space. It is important to remark
that only an interaction presenting the property of universality can be described by a
geometrization of spacetime. It is also important to mention that, in the eventual absence of
universality, the general relativity description of gravitation would break down.

On the other hand, like the other fundamental interactions of nature, gravitation can also
be described in terms of a gauge theory. In fact, the teleparallel equivalent
of general relativity, or teleparallel gravity for short,\cite{foot1} can be interpreted as a
gauge theory for the translation group. In this theory, instead of torsion, curvature is assumed
to vanish. The corresponding underlying spacetime is, in this case, a Weitzenb\"ock
spacetime. In spite of this fundamental difference, the two theories are found to yield
equivalent descriptions of the gravitational interaction.\cite{equiva} This means that
curvature and torsion are able to provide, each one, equivalent descriptions of the
gravitational interaction. Conceptual differences, however, show up. According to general
relativity, curvature is used to {\it geometrize} spacetime. Teleparallelism, on the other
hand, attributes gravitation to torsion, but in this case torsion accounts for gravitation
not by geometrizing the interaction, but by acting as a {\it force}. As a consequence, there
are no geodesics in the teleparallel equivalent of general relativity, but only force
equations quite analogous to the Lorentz force equation of electrodynamics.\cite{paper1} We
may then say that the gravitational interaction can be described in terms of curvature, as
is usually done in general relativity, or {\em alternatively} in terms of torsion, in which
case we have the so called teleparallel gravity. Whether gravitation requires a curved or a
torsioned spacetime---or equivalently, a Riemann\cite{foot2} or a Weitzenb\"ock spacetime
structure---turns out to be, at least classically, a matter of convention.

\subsection{Teleparallel Gravity and the Weak Equivalence Principle}

Rephrasing the above arguments, we can say that universality of free fall is the reason for
gravitation to present, in addition to the fundamental teleparallel gauge approach, the
equivalent geometric description of general relativity. In fact, in order to attribute
gravitation to curvature, it is essential that gravitation be universal, or equivalently,
that the {\em weak} equivalence principle, which establishes the equality of {\em inertial}
and {\em gravitational} masses, be true. Only under these circumstances is it possible to
assure that all spinless particles of nature, independently of their internal constitution,
feel gravitation the same and, for a given set of initial conditions, follow the same
trajectory---a geodesic of the underlying Riemannian spacetime.

Now, as is widely known, the electromagnetic interaction is not universal: there exists no
electromagnetic equivalence principle. Nevertheless, Maxwell's theory, a gauge theory for the
Abelian group $U(1)$, describes quite consistently the electromagnetic interaction. Relying
then on the fact that Maxwell's theory and teleparallel gravity are both Abelian gauge
theories, in which the equations of motion of test particles are not geodesics but force
equations, the question then arises whether the gauge approach of teleparallel gravity would
also be able to describe the gravitational interaction in the eventual lack of
universality---that is, of the weak equivalence principle. The answer to this question is
positive: teleparallel gravity does not require the validity of the equivalence principle to
describe the gravitational interaction.\cite{wep} In other words, whereas the geometrical
description of general relativity breaks down, the gauge description of teleparallel gravity
remains as a consistent theory in the absence of universality. In spite of the equivalence
with general relativity, therefore, teleparallel gravity seems to belong to a more general
class of theory. This is a very important issue because, even though the equivalence
principle has presently passed all experimental tests,\cite{exp} there are many controversies
related with its validity,\cite{synge} mainly at the quantum level.\cite{quantu}

\subsection{Teleparallel Coupling of the Fundamental Fields}

The gravitational coupling of the fundamental fields in teleparallel gravity is a very
controversial subject.\cite{hehl2}$^-$\cite{op2} The basic difficulty lies in the definition
of the spin connection, and consequently in the correct form of the gravitational coupling
prescription. Since no experimental data are available to help decide, from the different
possibilities, which is the correct one, the only one can do is to use consistency arguments
grounded in physical principles. One such possibility is to rely on the alluded equivalence
between general relativity and teleparallel gravity. According to this formulation, each one
of the fundamental fields of nature---scalar, spinor, and electromagnetic---are required to
couple to torsion in a such a way to preserve the equivalence between teleparallel gravity
and general relativity. When this approach is applied to these fields, as we are going to
see, a teleparallel spin connection can naturally be defined,\cite{tsc} which yields quite
consistent results.

\subsection{The Physics of Torsion Beyond Teleparallel Gravity}

As already discussed, in general relativity torsion is assumed to vanish from the very
beginning, whereas in teleparallel gravity curvature is assumed to vanish. In spite of this
fundamental difference, the two theories are found to yield equivalent descriptions of the
gravitational interaction. An immediate implication of this equivalence is that curvature
and torsion turns out to be simply alternative ways of describing the gravitational field,
and consequently related to the same degrees of freedom of gravity. This property is
corroborated by the fact that the symmetric matter energy-momentum tensor appears as source
in both theories: as the source of curvature in general relativity, and as the source of
torsion in teleparallel gravity.

On the other hand, theoretical speculations have since the early days of general relativity
discussed the necessity of including torsion, {\em in addition to curvature}, in the
description of the gravitational interaction.\cite{hammond} For example,  more general
gravity theories, like Einstein-Cartan\cite{ecar} and gauge theories for the
Poincar\'e\cite{kibble} and the affine groups,\cite{hcmn} consider curvature and torsion as
representing independent degrees of freedom. In these theories, differently from teleparallel
gravity, torsion should become relevant only when spins are important. This could be the case
either at the microscopic level or near a neutron star. According to these models, therefore,
since torsion represents additional degrees of freedom in relation to curvature, new physical
phenomena would be expected from its presence.

Now, the above described difference rises a conceptual question in relation to the actual
role played by torsion in the description of the gravitational interaction. In fact, the two
physical interpretations described above are clearly conflictive: if one is correct, the
other is necessarily wrong. This is a typical problem to be solved by experiment. However,
due to the weakness of the gravitational interaction, there are no available data on the
gravitational coupling of the fundamental particles. In addition, no one has ever reported
new gravitational phenomena near a neutron star, for example, where the effects of torsion
would be relevant according to those more general gravity theories. Therefore, also in this
general case, the only one can do in the search of a gravitational coupling prescription is
to use consistency arguments grounded in known physical principles. A possible way to proceed
is to remember that the general covariance principle---seen as an {\em active} version of the
strong equivalence principle---naturally defines a gravitational coupling prescription. We
can then use it to obtain the form of such prescription in the presence of curvature and
torsion. It should be remarked that this procedure is general, and has already been
consistently applied to obtain the coupling prescription in the specific case of teleparallel
gravity.\cite{mospe} 

\subsection{Purposes and Strategy}

The basic purpose of this paper is to critically review the role played by torsion in the
description of the gravitational interaction. We will proceed according to the following
scheme. In Section~\ref{notation}, we introduce the basic definitions and set the notations
we are going to use. In Section~\ref{telegra}, we present a comprehensive review of the
basic properties of teleparallel gravity. In particular, the role played by torsion in this
theory will be extensively discussed and clarified. In Section~\ref{gauge}, by requiring
compatibility with the strong equivalence principle, a gravitational coupling prescription
in the presence of curvature and torsion is obtained. This prescription, as we are going to
see, is found to be always equivalent with that of general relativity, a result that
reinforces the completeness of this theory, as well as the teleparallel point of view,
according to which torsion does not represent additional degrees of freedom for gravity, but
simply an alternative way of representing the gravitational field.\cite{newt1} An
application to the case of a spinning particle will be presented.\cite{newt2} Finally, a
discussion of the main points of the review will be made, and the consequences for
gravitation will be analyzed.

\section{Notations and Definitions}
\label{notation}

The geometrical setting of any gauge theory for gravitation is the tangent bundle, a natural
construction always present in spacetime. In fact, at each point of spacetime---the base
space of the bundle---there is always a tangent space attached to it---the fiber of the
bundle---on which the gauge group acts. We use the Greek alphabet
$(\mu, \nu, \rho, \dots = 0,1,2,3)$ to denote the holonomic indices related to spacetime, and
the Latin alphabet $(a,b,c, \dots = 0,1,2,3)$ to denote the anholonomic indices related to
the tangent space, assumed to be a Minkowski spacetime with the metric
$\eta_{ab}=\mathrm{diag}(+1,-1,-1,-1)$. The spacetime coordinates, therefore, will be denoted
by $x^\mu$, whereas the tangent space coordinates will be denoted by $x^a$. Since these
coordinates are functions of each other, the holonomic derivatives in these two spaces can be
identified by
\be
\partial_\mu = \left(\partial_{\mu} x^a\right) \;
\partial_a \quad \mathrm{and} \quad \partial_a =
\left(\partial_a x^{\mu}\right) \; \partial_\mu,
\label{1}
\ee where $\partial_{\mu} x^a$ is a trivial---that is, holonomic---tetrad, with $\partial_a
x^{\mu}$ its inverse.

From the geometrical point of view, a connection specifies how a vector field is
transported along a curve. In a local coordinate chart with basis vectors $\{e_\mu\} =
\{\partial_\mu\}$, the connection coefficients
$\Gamma^\lambda{}_{\nu\mu}$ are defined by
\be\label{twom}
\nabla_{e_\nu} e_\mu = e_\lambda \, \Gamma^\lambda{}_{\nu\mu}.
\ee
Once the action of $\nabla$ on the basis vectors is defined, one can compute
its action on any vector field $V^\rho$:
\be\label{threem}
\nabla_\mu V^\rho = \partial_\mu V^\rho +
\Gamma^\rho{}_{\nu \mu} \, V^\nu.
\ee
Now, given a nontrivial tetrad $h^a{}_\mu$, the spacetime and the tangent--space metrics
are related by
\be
g_{\mu\nu} = \eta_{ab} \, h^a{}_\mu \, h^b{}_\nu.
\label{gmn0}
\ee
Of course, as far as $e_\mu$ is a trivial tetrad, the metric $g_{\mu \nu} = \eta_{ab} \, e^a{}_\mu \,
e^b{}_\nu$ will be simply the Minkowski metric written in a different coordinate system. A connection
$\Gamma^\rho{}_{\lambda\mu}$ is said to be metric compatible if
\be \label{fourm}
\nabla_\lambda g_{\mu\nu} \equiv \partial_\lambda
g_{\mu\nu}-\Gamma^\rho{}_{\lambda\mu}g_{\rho\nu}-
\Gamma^\rho{}_{\lambda\nu}g_{\rho\mu} = 0.
\ee
On the other hand, a spin connection $A_\mu$ is a connection assuming values in the Lie
algebra of the Lorentz group,
\be
A_\mu = \onehalf \, A^{ab}{}_\mu \, S_{ab},
\ee
with $S_{ab}$ a representation of the Lorentz generators. Using the tetrad, a general
connection $\Gamma^{\rho}{}_{\nu \mu}$ can be related with the corresponding
spin connection $A^{a}{}_{b
\mu}$ through
\be
\Gamma^{\rho}{}_{\nu \mu} = h_{a}{}^{\rho} \partial_{\mu} h^{a}{}_{\nu} +
h_{a}{}^{\rho} A^{a}{}_{b \mu} h^{b}{}_{\nu}.
\label{geco}
\ee
The inverse relation is, consequently,
\be
A^{a}{}_{b \mu} =
h^{a}{}_{\nu} \partial_{\mu}  h_{b}{}^{\nu} +
h^{a}{}_{\nu} \Gamma^{\nu}{}_{\rho \mu} h_{b}{}^{\rho}.
\label{gsc}
\ee
Equations (\ref{geco}) and (\ref{gsc}) are simply different ways of expressing the property
that the total---that is, acting on both indices---derivative of the tetrad vanishes
identically:
\be
\partial_{\mu} h^{a}{}_{\nu} - \Gamma^{\rho}{}_{\nu \mu} h^{a}{}_{\rho} +
A^{a}{}_{b \mu} h^{b}{}_{\nu} = 0.
\label{todete}
\ee

In the present work, we will separate the notions of space and connections. From a formal
point of view, curvature and torsion are in fact properties of a connection.\cite{koba}
Strictly speaking, there is no such a thing as curvature or torsion of spacetime, but only
curvature or torsion of connections. This becomes evident if we remember that many different
connections are allowed to exist in the same spacetime.\cite{livro} Of course, when
restricted to the specific case of general relativity, universality of gravitation allows the
Levi--Civita connection to be interpreted as part of the spacetime definition as all
particles and fields feel this connection the same. However, when considering several
connections with different curvature and torsion, it seems far wiser and convenient to take
spacetime simply as a manifold, and connections (with their curvatures and torsions) as
additional structures.

The curvature and the torsion tensors of the connection $A^{a}{}_{b \mu}$ are defined
respectively by
\be
R^a{}_{b \nu \mu} = \partial_{\nu} A^{a}{}_{b \mu} -
\partial_{\mu} A^{a}{}_{b \nu} + A^a{}_{e \nu} A^e{}_{b \mu}
- A^a{}_{e \mu} A^e{}_{b \nu}
\ee
and
\be
T^a{}_{\nu \mu} = \partial_{\nu} h^{a}{}_{\mu} -
\partial_{\mu} h^{a}{}_{\nu} + A^a{}_{e \nu} h^e{}_{\mu}
- A^a{}_{e \mu} h^e{}_{\nu}.
\ee
Using the relation (\ref{gsc}), they can be expressed in a purely spacetime form, given by
\be
\label{sixbm}
R^\rho{}_{\lambda\nu\mu} \equiv h_a{}^\rho \, h^b{}_\lambda \, R^a{}_{b \nu \mu}
= \partial_\nu \Gamma^\rho{}_{\lambda \mu} -
\partial_\mu \Gamma^\rho{}_{\lambda \nu} +
\Gamma^\rho{}_{\eta \nu} \Gamma^\eta{}_{\lambda \mu} -
\Gamma^\rho{}_{\eta \mu} \Gamma^\eta{}_{\lambda \nu}
\ee
and
\be \label{sixam}
T^\rho{}_{\nu \mu} \equiv h_a{}^\rho \, T^a{}_{\nu \mu} =
\Gamma^\rho{}_{\mu\nu}-\Gamma^\rho{}_{\nu\mu}.
\ee

The connection coefficients can be conveniently decomposed according to
\be
\Gamma^\rho{}_{\mu\nu} = {\stackrel{\circ}{\Gamma}}{}^{\rho}{}_{\mu \nu} +
K^\rho{}_{\mu\nu},
\label{prela0}
\ee
where
\be
{\stackrel{\circ}{\Gamma}}{}^{\sigma}{}_{\mu \nu} = {\textstyle
\frac{1}{2}} g^{\sigma \rho} \left( \partial_{\mu} g_{\rho \nu} +
\partial_{\nu} g_{\rho \mu} - \partial_{\rho} g_{\mu \nu} \right)
\label{lci}
\ee
is the Levi--Civita connection of general relativity, and
\be
K^\rho{}_{\mu\nu} = {\textstyle
\frac{1}{2}} \left(T_\nu{}^\rho{}_\mu+T_\mu{}^\rho{}_\nu+
T^\rho{}_{\mu\nu}\right)
\label{contor}
\ee
is the contortion tensor. Using the relation (\ref{geco}), the decomposition (\ref{prela0})
can be rewritten in terms of the spin connections as
\be
A^c{}_{a\nu} = \Abol^c{}_{a\nu} + K^c{}_{a\nu},
\label{rela00}
\ee
where $\Abol^c{}_{a \nu}$ is the Ricci coefficient of rotation, the spin connection of
general relativity.

Teleparallel gravity, on the other hand, is characterized by the vanishing of the so called
Weitzenb\"ock spin connection: $\Aw^a{}_{b\mu}=0$. In this case, the relation (\ref{rela00})
assumes the form
\be
\Abol^c{}_{a\nu} = 0 - \Kw^c{}_{a\nu}.
\label{rela000}
\ee
Furthermore, from Eq.~(\ref{geco}) we see that the corresponding Weitzenb\"ock connection
has the form
\be
\Gammaw^{\rho}{}_{\nu \mu} = h_{a}{}^{\rho} \partial_{\mu} h^{a}{}_{\nu}.
\label{gecow}
\ee
In the remaining of the paper, all magnitudes related with general relativity will be
denoted with an over ``$\circ$'', whereas magnitudes related with teleparallel gravity
will be denoted with an over ``$\bullet$''.

Under a local Lorentz transformation $\Lambda^{a}{}_{b} \equiv
\Lambda^{a}{}_{b}(x)$, the tetrad changes according to $h'^{a}{}_{\mu} =
\Lambda^{a}{}_{b} \; h^{b}{}_{\mu}$, whereas the spin connection undergoes the
transformation
\be
A'^{a}{}_{b \mu} = \Lambda^{a}{}_{c} \, A^{c}{}_{d \mu} \,
\Lambda_{b}{}^{d} +
\Lambda^{a}{}_{c} \, \partial_{\mu} \Lambda_{b}{}^{c}.
\label{ltsc}
\ee
In the same way, it is easy to verify that $T^a{}_{\nu \mu}$ and $R^a{}_{b \nu \mu}$
transform covariantly under local Lorentz transformations:
\be
T'^{a}{}_{\nu \mu} = \Lambda^{a}{}_{b} \, T^b{}_{\nu \mu} \quad {\mathrm and} \quad
R'^{a}{}_{{b} \nu \mu} = \Lambda^{a}{}_{c} \,\Lambda_b{}^d \, R^c{}_{d \nu \mu}.
\ee
This means that $\Gamma^{\rho}{}_{\nu \mu}$, $T^\lambda{}_{\mu\nu}$ and
$R^\rho{}_{\lambda\nu\mu}$ are all invariant under a local Lorentz transformation.

A nontrivial tetrad field defines naturally a non-coordinate basis for vector fields and
their duals,
\be \label{ninem}
h_a = h_a{}^\mu \partial_\mu \quad \mathrm{and} \quad h^a = h^a{}_\mu dx^\mu.
\ee
This basis is clearly non-holonomic,
\be \label{tenm}
[h_c , h_d]=f^a{}_{cd} \, h_a,
\ee
with
\be
f^a{}_{cd} = h_c{}^\mu \, h_d{}^\nu (\partial_\nu
h^a{}_\mu - \partial_\mu h^a{}_\nu)
\ee
the coefficient of anholonomy. In this non-coordinate basis, and using the fact that the
last index of the spin connection is a tensor index,
\be
A^a{}_{bc} = A^a{}_{b \mu} \, h_c{}^\mu,
\ee
the curvature and torsion components are given respectively by\cite{abp1}
\be \label{13bm}
R^a{}_{bcd} = h_c A^a{}_{bd}-
h_d A^a{}_{bc} + A^a{}_{ec} \, A^a{}_{bd}
- A^a{}_{ed} \, A^e{}_{bc} + f^e{}_{cd} \, A^a{}_{be}
\ee
and
\be\label{13am}
T^a{}_{bc} = A^a{}_{cb} - A^a{}_{bc} - f^a{}_{bc}.
\ee

\section{Teleparallel Descriptions of Gravitation}
\label{telegra}

\subsection{Fundamentals of Teleparallel Gravity}

The notion of absolute parallelism (or teleparallelism) was introduced by Eins\-tein in the
late twenties, in his unsuccessful attempt to unify gravitation and
electromagnetism.\cite{sauer1} About three decades later, after the pioneering works by
Utiya\-ma\cite{utiyama} and Kibble,\cite{kibble} respectively on gauge theories for the
Lorentz and the Poincar\'e groups, there was a gravitationally-related revival of those
ideas,\cite{moller}$^-$\cite{haya} which since then have received considerable attention,
mainly in the context of gauge theories for gravitation,\cite{hehl}$^-$\cite{blago} of which
teleparallel gravity,\cite{hayshi}$^-$\cite{nes2} a gauge theory for the translation group,
is a particular case.

As a gauge theory for the translation group, the fundamental field of teleparallel
gravity is the gauge potential $B_{\mu}$, a field assuming values in the Lie algebra
of the translation group,
\be
B_{\mu} = B^{a}{}_{\mu} \, P_a,
\label{B}
\ee
where $P_{a} = \partial /\partial x^a$ are the translation generators, which satisfy
\be
\left[P_a , P_b \right] = 0.
\label{comu}
\ee
A gauge transformation is defined as a local (point dependent) translation of the
tangent-space coordinates,
\be
x^{a'} = x^a + \alpha^a,
\label{traxf}
\ee
with $\alpha^{a}\equiv\alpha^{a}(x^\mu)$ the corresponding infinitesimal parameters. In
terms of $P_a$, it can be written in the form
\begin{equation}
\delta x^{a} = \alpha^{b} P_{b} \, x^{a}.
\end{equation}

Let us consider now a general source field $\Psi\equiv\Psi(x^\mu)$. Its infinitesimal
gauge transformation does not depend on the spin character, and is given by
\be
\delta \Psi = \alpha^a P_a \Psi,
\label{trafi}
\ee
with $\delta \Psi$ standing for the functional change at the same $x^\mu$, which is the
relevant transformation for gauge theories. It is important to remark that the translation
generators are able to act on the argument of any source field because of the identifications
(\ref{1}). Using the general definition of covariant derivative\cite{livro}
\be
h_\mu = \partial_\mu + B^a{}_{\mu} \,
\frac{\delta }{\delta \alpha^a},
\label{cova}
\ee
the {\it translational} covariant derivative of $\Psi$ is found to be
\be
h_\mu \Psi = \partial_\mu \Psi + B^a{}_{\mu} \,
P_a \Psi.
\label{dfi1}
\ee
Equivalently, we can write\cite{paper1}
\be
h_\mu \Psi = h^{a}{}_{\mu} \; \partial_{a} \Psi,
\ee
where
\be
h^{a}{}_{\mu} = \partial_{\mu}x^{a} + B^{a}{}_{\mu} \equiv h_\mu x^a
\label{tetrada}
\ee
is a nontrivial---that is, anholonomic---tetrad field. As the generators $P_a=\partial_{a}$
are derivatives which act on the fields through their arguments, every source field in nature
will respond to their action, and consequently will couple to the translational gauge
potentials. In other words, every source field in nature will feel gravitation the same. This
is the origin of the concept of {\em universality} according to teleparallel gravity.

As usual in gauge theories, the field strength, denoted $\Tw^{a}{}_{\mu \nu}$, is
obtained from the commutation relation of covariant derivatives:
\begin{equation}
[h_{\mu}, h_{\nu}] \Psi = \Tw^{a}{}_{\mu \nu} P_a \Psi.
\label{commu1}
\end{equation}
We see from this expression that $\Tw^{a}{}_{\mu \nu}$ is also a field assuming values in
the Lie algebra of the translation group. As an easy calculation shows,
\begin{equation}
\Tw^{a}{}_{\mu \nu} = \partial_{\mu} B^{a}{}_{\nu}
- \partial_{\nu} B^{a}{}_{\mu} \equiv \partial_{\mu} h^{a}{}_{\nu}
- \partial_{\nu} h^{a}{}_{\mu}.
\label{core}
\end{equation}

Now, from the covariance of $h_{\mu} \Psi$, we obtain the transformation of
the gauge potentials:
\begin{equation}
B^{a^{\prime}}{}_{\mu} = B^{a}{}_{\mu} -
\partial_{\mu} \alpha^{a}.
\label{btrans}
\end{equation}
By using the transformations (\ref{traxf}) and (\ref{btrans}), the tetrad is found
to be gauge invariant:
\begin{equation}
h^{a'}{}_\mu = h^a{}_\mu.
\end{equation}
Consequently, as expected for an Abelian gauge theory, $\Tw^a{}_{\mu \nu}$ is also invariant
under a gauge transformation. We remark finally that, whereas the tangent space indices are
raised and lowered with the metric $\eta_{a b}$, the spacetime indices are raised and lowered
with the Riemannian metric $g_{\mu \nu}$, as given by Eq.~(\ref{gmn0}). It should be stressed
that, although representing the spacetime metric, $g_{\mu \nu}$ plays no dynamic role in the
teleparallel description of gravitation.

A nontrivial tetrad field induces on spacetime a teleparallel structure which is directly
related to the presence of the gravitational field. In fact, given a nontrivial tetrad, it is
possible to define the so called Weitzenb\"ock connection
\be
\Gammaw^{\rho}{}_{\mu \nu} = h_{a}{}^{\rho}\partial_{\nu}h^{a}{}_{\mu},
\label{carco}
\ee
which is a connection presenting torsion, but no curvature. As a natural consequence
of this definition, the Weitzenb\"ock covariant derivative of the tetrad field vanishes
identically:
\be
\nablaw_{\nu}h^a{}_{\mu} \equiv \partial_{\nu}h^a{}_{\mu} -
\Gammaw^{\rho}{}_{\mu \nu} \, h^a{}_{\rho} = 0.
\label{cacd}
\ee
This is the absolute parallelism condition. The torsion of the Weitzenb\"ock
connection is
\be
\Tw^{\rho}{}_{\mu \nu} = \Gammaw^{\rho}{}_{\nu \mu} -
\Gammaw^{\rho}{}_{\mu \nu},
\label{tor}
\ee
from which we see that the gravitational field strength is nothing but torsion
written in the tetrad basis:
\be
\Tw^{a}{}_{\mu \nu} = h^{a}{}_{\rho} \, \Tw^{\rho}{}_{\mu \nu}.
\label{fht}
\ee
In terms of $\Tw^{\rho}{}_{\mu \nu}$, the commutation relation (\ref{commu1}) assumes the
form
\be
\left[h_\mu, h_\nu \right] = \Tw^\rho{}_{\mu \nu} \; h_\rho,
\label{comuta2}
\ee
from where we see that torsion plays also the role of the nonholonomy of the
translational gauge covariant derivative.

A nontrivial tetrad field can also be used to define a torsionless linear
connection $\Gammabol^{\rho}{}_{\mu \nu}$, the Levi--Civita connection of the
metric (\ref{gmn0}), given by Eq.~(\ref{lci}).
The Weitzenb\"ock and the Levi--Civita connections are related by
\be
\Gammaw^{\rho}{}_{\mu \nu} =
{\stackrel{\circ}{\Gamma}}{}^{\rho} {}_{\mu \nu} +
\Kw^{\rho}{}_{\mu \nu},
\label{rela0}
\ee
where
\be
\Kw^{\rho}{}_{\mu \nu} = {\textstyle \frac{1}{2}} \left( \Tw_{\mu}{}^{\rho}{}_{\nu}
+ \Tw_{\nu}{}^{\rho}{}_{\mu} - \Tw^{\rho}{}_{\mu \nu} \right)
\ee
is the contortion of the Weitzenb\"ock torsion.

As already remarked, the curvature of the Weitzenb\"ock connection vanishes
identically:
\be
\Rw^{\rho}{}_{\theta \mu \nu} = \partial_\mu
\Gammaw^{\rho}{}_{\theta \nu} -\partial_\nu
\Gammaw^{\rho}{}_{\theta \mu} + \Gammaw^{\rho}{}_{\sigma
\mu} \; \Gammaw^{\sigma}{}_{\theta \nu} - \Gammaw^{\rho}{}_{\sigma
\nu} \; \Gammaw^{\sigma}{}_{\theta \mu} \equiv 0.
\label{wr}
\ee
Substituting $\Gammaw^{\rho}{}_{\mu \nu}$ as given by Eq.~(\ref{rela0}), we get
\be
\Rw^{\rho}{}_{\theta \mu \nu} =
{\stackrel{\circ}{R}}{}^{\rho}{}_{\theta \mu \nu} +
\Qw^{\rho}{}_{\theta \mu \nu} \equiv 0,
\label{relar}
\ee
where $\Rbol^{\theta}{}_{\rho \mu \nu}$ is the curvature of the Levi--Civita connection,
and
\be
\Qw^{\rho} {}_{\theta \mu \nu} = \dw_{\mu}{}\Kw^{\rho}{}_{\theta \nu} -
\dw_{\nu}{}\Kw^{\rho}{}_{\theta \mu} + \Kw^{\sigma}{}_{\theta \nu}
\; \Kw^{\rho}{}_{\sigma \mu} - \Kw^{\sigma}{}_{\theta \mu} \;
\Kw^{\rho}{}_{\sigma \nu}
\label{qdk}
\ee
is a tensor written in terms of the Weitzenb\"ock connection only. Here, $\dw_\mu$ is
the {\it teleparallel covariant derivative}, with a connection term for each index of
$\Kw^{\rho}{}_{\theta \nu}$. Acting on a spacetime vector $V^\mu$, for  example, its
explicit form is
\be
\dw_\rho V^\mu \equiv \partial_\rho V^\mu +
\left( \Gammaw^\mu{}_{\lambda \rho} - \Kw^\mu{}_{\lambda \rho} \right) V^\lambda.
\label{tcd}
\ee
Owing to the relation (\ref{rela0}), we see that it is nothing but the Levi--Civita
covariant derivative of general relativity rephrased in terms of the Weitzenb\"ock
connection.

Equation (\ref{relar}) has an interesting interpretation: the contribution
${\stackrel{\circ}{R}}{}^{\rho}{}_{\theta \mu \nu}$ coming from the
Levi--Civita connection compensates exactly the contribution
$Q^{\rho}{}_{\theta \mu \nu}$ coming from the Weitzenb\"ock connection, yielding
an identically zero curvature tensor $\Rw^{\rho}{}_{\theta \mu \nu}$. This is a
constraint satisfied by the Levi--Civita and Weitzenb\"ock connections, and is the
fulcrum of the equivalence between the Riemannian and the teleparallel descriptions
of gravitation.

\subsection{Spin Connection and Coupling Prescription}

\subsubsection{General Relativity Spin Connection}

The interaction of a general matter field with gravitation can be obtained through the
application of the so called minimal coupling prescription, according to which all
ordinary derivatives must be replaced by covariant derivatives. Because they are used
in the construction of these covariant derivatives, gauge connections (or potentials,
in physical terminology) are the most important personages in the description of an
interaction. The relevant spin connection of general relativity is the so called Ricci
coefficient of rotation $\stackrel{~\circ}{A}_{\mu}$, a torsionless connection assuming
values in the Lie algebra of the Lorentz group:
\be
\stackrel{~\circ}{A}_{\mu} = \onehalf \,
{\stackrel{~\circ}{A}}{}^{a b}{}_{\mu} \, S_{a b},
\label{spinco}
\ee
where $S_{a b}$ is an element of the Lorentz Lie algebra written in some appropriate
representation. The minimal coupling prescription in general relativity, therefore,
amounts to replace
\be
\partial_a \rightarrow {\stackrel{\circ}{\mathcal D}}_a = h_a{}^\mu \,
{\stackrel{\circ}{\mathcal D}}_\mu,
\ee
where
\be
{\stackrel{\circ}{\mathcal D}}_\mu = \partial_\mu -  \frac{i}{2} \,
{\stackrel{~\circ}{A}}{}^{a b}{}_{\mu} \, S_{a b}
\label{fi}
\ee
is the Fock-Ivanenko covariant derivative.\cite{fi} It is important to remark that the
spin connection ${\stackrel{~\circ}{A}}{}^{a b}{}_{\mu}$ is not an independent field. In
fact, in terms of the Levi-Civita connection, it is written as\cite{dirac}
\be
{\stackrel{~\circ}{A}}{}^{a}{}_{b \mu} = h^{a}{}_{\rho} \,
{\stackrel{\circ}{\Gamma}}{}^{\rho}{}_{\mu \nu} \, h_{b}{}^{\mu} +
h^{a}{}_{\rho} \, \partial_\nu h_{b}{}^{\rho} \equiv
h^{a}{}_{\rho} \stackrel{\circ}{\nabla}_\nu h_{b}{}^{\rho},
\label{gra}
\ee
from where we see that it is totally determined by the tetrad---or equivalently, by the
metric. This means that, in general relativity, the local Lorentz symmetry is not
dynamical (gauged), but essentially a kinematic symmetry.

Now, as is well known, a tetrad field can be used to transform {\em Lorentz} into {\em
spacetime} tensors, and {\em vice-versa}. For example, a Lorentz vector field $V^a$ is
related to the corresponding spacetime vector $V^\mu$ through
\be
V^a = h^a{}_{\mu} V^\mu.
\label{tatmu}
\ee
As a consequence, the covariant derivative (\ref{fi}) of a general {\em Lorentz}
tensor field reduces to the usual Levi-Civita covariant derivative of the corresponding
{\em spacetime} tensor. For example, take the vector field $V^a$ for which the
appropriate Lorentz generator is\cite{ramond}
\be
(S_{ab})^c{}_d = i \left( \delta^c{}_a \, \eta_{bd} - \delta^c{}_b \, \eta_{ad}
\right).
\label{vecre}
\ee
It is then an easy task to verify that
\be
{\stackrel{\circ}{\mathcal D}}_\mu V^a = h^{a}{}_{\rho} \,
{\stackrel{\circ}{\nabla}}_{\mu} V^\rho \; .
\ee

However, in the case of half-integer spin fields, the situation is completely different.
In fact, as is well known, there exists no spacetime representation for spinor
fields.\cite{veltman} This means that no Levi-Civita covariant derivative can be defined for
these fields. Thus, the only possible form for the covariant derivative of a Dirac spinor
$\psi$, for example, is that given in terms of the spin connection,
\be
{\stackrel{\circ}{\mathcal D}}_\mu \psi = \partial_\mu \psi - \frac{i}{2} \,
{\stackrel{~\circ}{A}}{}^{a b}{}_{\mu} \, S_{a b} \,\psi,
\label{di}
\ee
where
\be
S_{ab} = \frac{i}{4} [\gamma_a, \gamma_b]
\label{spinlore}
\ee
is the Lorentz spin-1/2 generator, with $\gamma_a$ the Dirac matrices. We may say, therefore,
that the Fock-Ivanenko derivative ${\stackrel{\circ}{\mathcal D}}_\mu$ is more fundamental
than the Levi-Civita covariant derivative ${\stackrel{\circ}{\nabla}}_\mu$ in the sense that
it is able to describe, through the minimal coupling prescription, the gravitational
interaction of both tensor and spinor fields.

\subsubsection{Teleparallel Spin Connection}

In order to obtain the teleparallel version of the minimal coupling prescription, it
is necessary to find first the correct teleparallel spin connection. Inspired by the
definition (\ref{gra}), which gives the general relativity spin connection, it is usual
to start by making the following attempt,
\be
\Aw^{a}{}_{b \mu} = h^{a}{}_{\rho} \, \Gammaw^{\rho}{}_{\nu \mu} \,
h_{b}{}^{\mu} + h^{a}{}_{\rho} \, \partial_\mu h_{b}{}^{\rho} \equiv
h^{a}{}_{\rho} \, \nablaw_\mu h_{b}{}^{\rho}.
\label{telespincon}
\ee
However, as a consequence of the absolute parallelism condition (\ref{cacd}), we see that
$\Aw^{a}{}_{b \mu}=0$. This does not mean that in teleparallel gravity the {\it dynamical}
spin connection, that is, the spin connection defining the minimal coupling prescription,
vanishes. In fact, notice that, due to the affine character of connection space, there exists
infinitely more possibilities.

Let us then adopt a different procedure to look for the teleparallel spin connection.
Our basic guideline will be to find a coupling prescription which results equivalent to
the coupling prescription of general relativity. This can be achieved by taking Eq.
(\ref{rela0}) and rewriting it in the tetrad basis. By using the transformation
properties (\ref{gra}) and (\ref{telespincon}), we get
\be
{\stackrel{\circ}{A}}{}^{a}{}_{b \mu} = - \Kw^{a}{}_{b \mu} + 0,
\label{tsc}
\ee
where
\be
\Kw^{a}{}_{b \mu} = h^a{}_\rho \, \Kw^{\rho}{}_{\nu \mu} h_b{}^\nu,
\label{kfk}
\ee
and where we have already used that $\Aw^{a}{}_{b \mu}=0$. Notice that the {\em zero
connection} appearing in Eq.~(\ref{tsc}) is crucial in the sense that it is the
responsible for making the right hand-side a true connection. Therefore, based on these
considerations, we can say that the teleparallel spin connection is given by {\em minus}
the contortion tensor plus a zero-connection:\cite{tsc}
\be
\Omegaw^{a}{}_{b \mu} = - \Kw^{a}{}_{b \mu} + 0.
\ee

Like any connection assuming values in the Lie algebra of the Lorentz group,
\be
\Omegaw_{\mu} = \onehalf \, \Omegaw^{a}{}_{b \mu} \, S_a{}^b,
\ee
$\Omegaw^{a}{}_{b \mu}$ is anti-symmetric in the first two indices. Furthermore, under an
infinitesimal local Lorentz transformation with parameters $\epsilon^a{}_b \equiv
\epsilon^a{}_b(x^\mu)$, it changes according to
\be
\delta \Omegaw^{a}{}_{b \mu} = - \Dw_\mu \epsilon^a{}_b,
\label{ktrans1}
\ee
where
\be
\Dw_\mu \epsilon^a{}_b = \partial_\mu \epsilon^a{}_b +
 \Omegaw^{a}{}_{c \mu} \, \epsilon^c{}_b -
 \Omegaw^{c}{}_{b \mu} \, \epsilon^a{}_c
\label{ktrans2}
\ee
is the covariant derivative with $\Omegaw^{a}{}_{b \mu}$ as the connection. Equation
(\ref{ktrans1}) is the standard gauge potential transformation of non-Abelian gauge
theories. We see in this way that, in fact, $\Omegaw^{a}{}_{b \mu}$ plays
the role of the spin connection in teleparallel gravity. Accordingly, the
teleparallel Fock-Ivanenko derivative operator is to be written in the form
\be
\Dw_\mu = \partial_\mu - \frac{i}{2} \,
\Omegaw^{a}{}_{b \mu} \, S_a{}^b.
\label{tfi}
\ee
Notice that (\ref{ktrans2}) is a particular case of this covariant derivative,
obtained by taking $S_a{}^b$ as the spin-2 representation of the Lorentz
generators.\cite{ramond} The minimal coupling prescription of teleparallel gravity,
therefore, can be written in the form
\be
\partial_a \rightarrow \Dw_a = h_a{}^\mu \, \Dw_\mu,
\label{tcp}
\ee
with $\Dw_\mu$ the teleparallel Fock-Ivanenko derivative (\ref{tfi}).

The covariant derivative (\ref{tfi}) presents all necessary properties to be
considered as yielding the fundamental coupling prescription in teleparallel gravity.
For example, it transforms covariantly under local Lorentz transformations:
\be
\Dw^\prime_\mu = U \Dw_\mu U^{-1}.
\ee
Another important property is that the teleparallel coupling prescription defined by
the covariant derivative (\ref{tfi}) turns out to be completely equivalent with the
usual minimal coupling prescription of general relativity. Actually, it is just what is
needed so that it turns out to be the minimal coupling prescription of general
relativity rephrased in terms of magnitudes of the teleparallel structure. Analogously
to general relativity, the teleparallel Fock-Ivanenko covariant derivative (\ref{tfi})
is the only one available for spinor fields in teleparallel gravity. For tensor fields,
on the other hand, there is also a spacetime covariant derivative which acts in the
corresponding spacetime tensors. As an example, let us consider again a Lorentz vector
field $V^a$. By using the vector generator (\ref{vecre}), it is an easy task to show that
\be
\Dw_\mu V^a = h^{a}{}_{\rho} \, \dw_{\mu} V^\rho,
\ee
where $\dw_{\mu}$ is the teleparallel covariant derivative (\ref{tcd}).

\subsubsection{Further Remarks}

Due to the fact that the Weitzenb\"ock spin connection $\Aw$ vanishes when written in the
tetrad basis, it is usually asserted that, for spinor fields, when written in the form
\be
\Dw_\mu = \partial_\mu - \frac{i}{2} \,
\Aw^{a}{}_{b \mu} \, S_a{}^b,
\label{wtfi}
\ee
the teleparallel Fock-Ivanenko derivative coincides with the ordinary
derivative:\cite{hayshi}
\be
\Dw_\mu = \partial_\mu
\ee
However, there are several problems associated to this coupling prescription. First, it
is not compatible with the coupling prescription of general relativity, which is somehow
at variance with the equivalence between the corresponding gravitational Lagrangians.
Second, it results different to apply this coupling prescription in the Lagrangian or
in the field equation, which is a rather strange result. Summing up, we could say that
there is no compelling arguments supporting the choice of $\Aw^a{}_{b
\mu}$ as the spin connection of teleparallel gravity.

On the other hand, several arguments favor the conclusion that the spin connection of
teleparallel gravity is in fact given by
\be
\Omegaw^{a}{}_{b \mu} = - \Kw^{a}{}_{b \mu} + 0,
\label{tsc2}
\ee
which leads to the coupling prescription (\ref{tfi}), or equivalently, to
\be
\Dw_\mu = \partial_\mu + \frac{i}{2} \,
\Kw^{a}{}_{b \mu} \, S_{a}{}^{b},
\label{tfibis}
\ee
where we have dropped the {\em zero connection} for simplicity of notation. First, it
is covariant under local Lorentz transformations. Second, in contrast to the coupling
prescription (\ref{wtfi}), it results completely equivalent to apply the minimal
coupling prescription in the Lagrangian or in the field equation. And third, it is
self-consistent, and agrees with general relativity. For example, it is well know that,
in the context of general relativity, the {\it total} covariant derivative of the
tetrad field vanishes,
\be
\partial_\nu h_{b}{}^{\rho} +
{\stackrel{\circ}{\Gamma}}{}^{\rho}{}_{\mu \nu} \, h_{b}{}^{\mu} -
\Abol^{a}{}_{b \nu} \, h_{a}{}^{\rho} = 0,
\label{rdh}
\ee
which is actually the same as (\ref{gra}). The teleparallel version of this expression
can be obtained by substituting ${\stackrel{\circ}{\Gamma}}{}^{\rho}{}_{\mu \nu}$ and
${\stackrel{~\circ}{A}}{}^{a}{}_{b \nu}$ by their teleparallel counterparts. Using
Eqs.~(\ref{rela0}) and (\ref{tsc}), one gets
\be
\partial_\nu h_{b}{}^{\rho} +
\left( \Gammaw^{\rho}{}_{\mu \nu} - \Kw^{\rho}{}_{\mu \nu} \right) h_b{}^\mu +
\Kw^{a}{}_{b \nu} \, h_{a}{}^{\rho} = 0.
\label{tdh}
\ee
However, the contortion terms cancel out, yielding the absolute parallelism condition
\be
\partial_\nu h_{b}{}^{\rho} +
\Gammaw^{\rho}{}_{\mu \nu} h_b{}^\mu = 0,
\label{apc}
\ee
which is the fundamental equation of teleparallel gravity. This shows the consistency
of identifying the teleparallel spin connection as {\em minus} the contortion tensor
plus the zero connection. Finally, as the coupling prescription (\ref{tfibis}) is
covariant under local Lorentz transformation and is equivalent with the minimal coupling
prescription of general relativity, teleparallel gravity with this coupling
prescription turns out to be completely equivalent to general relativity, even in the
presence of spinor fields.

\subsection{Application to the Fundamental Fields}

\subsubsection{Scalar Field}

According to the Einstein--Cartan models, only a spin distribution could produce or feel
torsion.\cite{hamm1} A scalar field, for example, should be able to feel curvature
only.\cite{hehl2} However, since from the teleparallel point of view the interaction of
gravitation with any field can be described alternatively in terms of curvature or torsion,
and since a scalar field is known to couple to curvature, it might also couple to
torsion.\cite{paper2} To see that, let us consider the Lagrangian of a scalar field $\phi$,
which in a Minkowski spacetime is given by
\begin{equation}
{\mathcal L}_{\phi} = \onehalf \left[ \eta^{a b} \; \partial_a \phi
\; \partial_b \phi - \mu^2 \phi^2 \right],
\label{fsf}
\end{equation}
where $\mu = {m c}/{\hbar}$. The corresponding field equation is the Klein--Gordon equation
\begin{equation}
\partial_a \partial^a \phi + \mu^2 \phi = 0.
\label{kge1}
\end{equation}
The coupling with gravitation is obtained by applying the teleparallel
coupling prescription
\begin{equation}
\partial_a \rightarrow \Dw_a \equiv h_a{}^\mu \, \Dw_\mu =
h_a{}^\mu \left( \partial_\mu + \frac{i}{2} \, \Kw^{a b}{}_{\mu} \, S_{a b} \right)
\label{fulltcp}
\end{equation}
to the free Lagrangian (\ref{fsf}). The result is
\begin{equation}
{\mathcal L}_{\phi} = \frac{h}{2} \left[ \eta^{a b} \, \Dw_a \phi \;
\Dw_b \phi - \mu^2 \phi^2 \right],
\label{lweitz1}
\end{equation}
or equivalently,
\begin{equation}
{\mathcal L}_{\phi} = \frac{h}{2} \left[ g^{\mu \nu} \, \Dw_\mu \phi \;
\Dw_\nu \phi - \mu^2 \phi^2 \right],
\label{lweitz2}
\end{equation}
where $h = \det(h^a{}_\mu)$. In the specific case of a scalar field, $S_{ab} \phi
= 0$, and consequently
\begin{equation}
\Dw_\mu \phi = \partial_\mu \phi.
\label{da}
\end{equation}
Using the identity
\begin{equation}
\partial_\mu h = h \, h_a{}^{\rho} \, \partial_\mu h^a{}_\rho
\equiv h \, \Gammaw^{\rho}{}_{\rho \mu},
\end{equation}
it is easy to show that the corresponding field equation is
\begin{equation}
{\qedw} \, \phi + \mu^2 \phi = 0,
\label{klein1}
\end{equation}
where
\begin{equation}
{\qedw} \; \phi = h^{-1} \; \partial_\rho \left(h \; \partial^\rho \phi \right) \equiv
\partial_\mu \partial^\mu \phi +
\Gammaw^{\mu}{}_{\mu \rho} \, \partial^\rho \phi
 \label{telb1}
\end{equation}
is the teleparallel version of the Laplace--Beltrami operator. Because
$\Gammaw^{\rho}{}_{\rho \mu}$ is not symmetric in the last two indices, the above
expression is not the Weitzenb\"ock covariant divergence of $\partial^\mu \phi$. From
Eq.~(\ref{tor}), however, we can write
\begin{equation}
\Gammaw^{\mu}{}_{\mu \rho} = \Gammaw^{\mu}{}_{\rho \mu} + \Tw^{\mu}{}_{\rho \mu},
\end{equation}
and the expression for ${\qedw} \, \phi$ may be rewritten in the form
\begin{equation}
{\qedw} \, \phi = \left({\nablaw}_\rho + \Tw^{\mu}{}_{\rho \mu} \right)
\partial^\rho \phi,
\label{telb2}
\end{equation}
from where we see that the scalar field is able to couple to torsion through the derivative
$\partial^\rho \phi$. Making use of the identity
\begin{equation}
\Tw^{\rho}{}_{\mu \rho} = - \Kw^{\rho}{}_{\mu \rho},
\end{equation}
the teleparallel version of the Klein--Gordon equation turns out to be\cite{paper2}
\begin{equation}
\dw_{\mu} \partial^\mu \phi + \mu^2 \phi = 0,
\label{klein3}
\end{equation}
with $\dw_\mu$ the teleparallel covariant derivative (\ref{tcd}).

\subsubsection{Dirac Spinor Field}

In teleparallel gravity, the coupling of spinor fields with gravitation is a highly
controversial subject.\cite{hehl2}$^-$\cite{op2}
The reason for this is that in teleparallel gravity the dynamical spin connection---that
is, the connection that describes the interaction of a spinor field with gravitation---is
assumed to vanish.\cite{hayshi} However, as we are going to see, if instead of zero the
connection (\ref{tsc}) is considered as the teleparallel spin connection, teleparallel
gravity becomes consistent and fully equivalent with general relativity, even in the presence
of spinor fields.

In Minkowski spacetime, the spinor field Lagrangian is
\begin{equation}
{\mathcal L}_\psi = \frac{i c \hbar}{2} \,
\left( \bar{\psi} \, \gamma^a \partial_a \psi -
\partial_a \bar{\psi} \gamma^a \, \psi \right) - m c^2 \, \bar{\psi} \psi.
\label{fspinl}
\end{equation}
The corresponding field equation is the free Dirac equation
\begin{equation}
i \hbar \gamma^a \, \partial_a \psi - m c \, \psi = 0.
\end{equation}
The gravitationally coupled Dirac Lagrangian is obtained
by applying the teleparallel coupling prescription (\ref{fulltcp}), with $S_{ab}$ the spinor
representation (\ref{spinlore}). The result is
\begin{equation}
{\mathcal L}_\psi = h \left[ \frac{i c \hbar}{2} \,
\left( \bar{\psi} \gamma^\mu \, \Dw_\mu \psi -
\Dw^*_\mu \bar{\psi} \, \gamma^\mu \psi \right) - m \, c^2 \, \bar{\psi} \psi
\right],
\label{tspinl}
\end{equation}
where $\gamma^\mu \equiv \gamma^\mu(x) = \gamma^a \, h_a{}^\mu$.
Using the identity $\Dw_\mu (h \gamma^\mu) = 0$, the teleparallel version of the
coupled Dirac equation is found to be
\begin{equation}
i \hbar \gamma^\mu \Dw_\mu \psi - m c \, \psi = 0.
\label{tcde}
\end{equation}
Comparing the Fock--Ivanenko derivatives (\ref{fi}) and (\ref{tfibis}), we see that, whereas
in general relativity the Dirac spinor couples to the Ricci coefficient of rotation
$\Abol_\mu$, in teleparallel gravity it couples to the contortion tensor $\Kw_\mu$.

\subsubsection{Electromagnetic Field}

In the usual framework of torsion gravity, it is a commonplace to assert that the
electromagnetic field cannot be coupled to torsion in order to preserve the local gauge
invariance of Maxwell's theory. Equivalently, one can say that, in the presence of torsion,
the requirement of gauge invariance precludes the existence of a gravitational minimal
coupling prescription for the electromagnetic field.\cite{hehl2} To circumvent this problem,
it is usually postulated that the electromagnetic field can neither produce nor feel
torsion.\cite{sha} In other words, torsion is assumed to be irrelevant to the Maxwell's
equations.\cite{bdt} This ``solution'' to the problem of the interaction of torsion with the
electromagnetic field is far from reasonable. A far more consistent solution is achieved by
observing that, if the electromagnetic field couples to curvature, the equivalence between
general relativity and teleparallel gravity implies necessarily that it must also couple to
torsion. In addition, it should be remarked that the above postulate is not valid at a
microscopic level since, from a quantum point of view, one may always expect an interaction
between photons and torsion.\cite{sasi} The reason for this is that a photon, perturbatively
speaking, can virtually disintegrate into an electron--positron pair, and as these particles
are massive fermions which couple to torsion, the photon must necessarily feel the presence
of torsion. Consequently, even not interacting directly with torsion, the photon field does
feel torsion through the virtual pair produced by the vacuum polarization. Moreover, as all
macroscopic phenomena must necessarily have an interpretation based on an average of
microscopic phenomena, and taking into account the strictly attractive character of
gravitation which eliminates the possibility of a vanishing average, the above hypothesis
seems to lead to a contradiction as no interaction is postulated to exist at the macroscopic
level. As we are going to see, in spite of the controversies,\cite{fhyo} provided the
teleparallel spin connection be properly chosen, the electromagnetic field can consistently
couple to torsion in teleparallel gravity.

In Minkowski spacetime, the electromagnetic field is described by the
Lagran\-gian density
\begin{equation}
{\mathcal L}_{em} = - \frac{1}{4} \, F_{ab} F^{ab},
\label{l0}
\end{equation}
where
\begin{equation}
F_{ab} = \partial_a A_b - \partial_b A_a
\label{f0}
\end{equation}
is the electromagnetic field-strength. The corresponding field equation is
\begin{equation}
\partial_a F^{ab} = 0,
\label{max0}
\end{equation}
which along with the Bianchi identity
\begin{equation}
\partial_a F_{bc} + \partial_c F_{ab} + \partial_b F_{ca} = 0,
\label{bian0}
\end{equation}
constitutes Maxwell's equations.  In the Lorentz gauge
$\partial_a A^a = 0$, the field equation (\ref{max0}) acquires the form
\begin{equation}
\partial_c \partial^c A^a = 0.
\label{amax0}
\end{equation}

Let us obtain now, by applying the coupling prescription (\ref{fulltcp}), Maxwell's
equation in teleparallel gravity.\cite{vector} In the specific case of the electromagnetic
vector field, the Lorentz generators $S_{ab}$ are written in the vector representation
(\ref{vecre}), and the Fock--Ivanenko derivative assumes the form
\begin{equation}
\Dw_\mu A^c = \partial_\mu A^c - \Kw^c{}_{d \mu} \, A^d.
\label{dac}
\end{equation}
To obtain the corresponding covariant derivative of the
spacetime vector field $A^\nu$, we substitute $A^d=h^{d}{}_{\nu}
A^\nu$ in the right-hand side.  The result is
\begin{equation}
\Dw_\mu A^c = h^c{}_{\rho} \dw_\mu A^\rho,
\label{dac2}
\end{equation}
with
\be
\dw_\mu A^\rho = \partial_\mu A^\rho + \left(\Gammaw^{\rho}{}_{\nu \mu} -
\Kw^{\rho}{}_{\nu \mu} \right) A^\nu
\label{telcode}
\end{equation}
the teleparallel covariant derivative.  This means that, for the specific case of a vector
field, the teleparallel version of the minimal coupling prescription (\ref{fulltcp}) can
alternatively be stated as
\begin{equation}
\partial_a A_b \rightarrow h_a{}^\mu h_b{}^\rho \dw_\mu A_\rho.
\label{dmc2}
\end{equation}
As a consequence, the gravitationally coupled Maxwell Lagrangian in teleparallel gravity
can be written as
\begin{equation} {\cal L}_{em} = - \frac{h}{4} \, F_{\mu \nu}
F^{\mu \nu},
\label{l2}
\end{equation}
where
\begin{equation}
F_{\mu \nu} = \dw_\mu A_\nu - \dw_\nu A_\mu.
\label{f2}
\end{equation}
Using the explicit form of $\dw_\mu$, and the definitions of torsion and contortion
tensors, it is easy to verify that
\begin{equation}
F_{\mu \nu} = \partial_\mu A_\nu - \partial_\nu A_\mu.
\label{f3}
\end{equation}
We notice in passing that this tensor is invariant under the $U(1)$ electromagnetic gauge
transformations. The corresponding field equation is
\begin{equation}
\dw_\mu F^{\mu \nu} = 0,
\end{equation}
which yields the first pair of Maxwell's equation in teleparallel gravity. Assuming the
teleparallel Lorentz gauge $\dw_\mu A^\mu = 0$, and using the commutation relation
\begin{equation}
\left[ \dw_\mu,
\dw_\nu \right] A^\mu = - \Qw_{\mu \nu} \, A^\mu,
\end{equation}
where $\Qw_{\mu \nu} = \Qw^{\rho}{}_{\mu \rho \nu}$, with $\Qw^{\rho}{}_{\mu \sigma \nu}$
given by Eq.~(\ref{qdk}), we obtain
\begin{equation}
\dw_{\mu} \dw^{\mu} A_{\nu}
+ \Qw^{\mu}{}_{\nu} A_{\mu} = 0.
\label{amax2}
\end{equation}
This is the teleparallel version of the first pair of Maxwell's equation. On the other
hand, by using the same coupling prescription in the Bianchi identity (\ref{bian0}), the
teleparallel version of the second pair of Maxwell's equation is found to be
\begin{equation} %
    \partial_\mu F_{\nu \sigma} +
\partial_\sigma F_{\mu \nu} + \partial_\nu F_{\sigma \mu} = 0.
\label{bian2}
\end{equation}

Summing up: in the context of the teleparallel equivalent of general relativity, the
electromagnetic field is able to couple to torsion, and that this coupling does not violate
the $U(1)$ gauge invariance of Maxwell's theory.  Furthermore, using the relation
(\ref{rela0}), it is easy to verify that the teleparallel version of Maxwell's equations,
which are equations written in terms of the Weitzenb\"ock connection only, are completely
equivalent with the usual Maxwell's equations in a Riemannian background, which are
equations written in terms of the Levi--Civita connection only. We can then say that
teleparallel gravity is able to provide a consistent description of the interaction of
torsion with the electromagnetic field.\cite{vector}

\subsection{Lagrangian and Field Equations}

The Lagrangian of the teleparallel equivalent of general relativity is
\be
\Lw = \frac{h}{2 k^2} \; \left[\frac{1}{4} \;
\Tw^\rho{}_{\mu \nu} \; \Tw_\rho{}^{\mu \nu} + \frac{1}{2} \;
\Tw^\rho{}_{\mu \nu} \; \Tw^{\nu \mu} {}_\rho - \Tw_{\rho \mu}{}^{\rho}
\; \Tw^{\nu \mu}{}_\nu \right],
\label{lagr3}
\ee
where $k^2 = 8 \pi G/c^{4}$ and $h = {\rm det}(h^{a}{}_{\mu})$. The first term
corresponds to the usual Lagrangian of gauge theories. In the gravitational case,
however, owing to the presence of a tetrad field, algebra and spacetime indices can now be
changed into each other, and in consequence new contractions turn out to be possible. It
is exactly this possibility that gives rise to the other two terms of the above
Lagrangian. If we define the tensor
\begin{equation}
\Sw^{\rho\mu\nu} = - \Sw^{\rho\nu\mu} =
\left[ \Kw^{\mu\nu\rho} - g^{\rho\nu}\,\Tw^{\sigma\mu}{}_{\sigma}
+ g^{\rho\mu}\,\Tw^{\sigma\nu}{}_{\sigma} \right],
\label{S}
\end{equation}
usually called superpotential,\cite{mollersbook} the teleparallel Lagrangian (\ref{lagr3})
can be rewritten in the form\cite{maluf}
\begin{equation}
\Lw =
\frac{h}{4 k^2} \, \Tw_{\rho\mu\nu} \, \Sw^{\rho\mu\nu}.
\label{gala}
\end{equation}
Using the identity
\be
T^\mu{}_{\mu \rho} = K^\mu{}_{\rho \mu},
\ee
which follows from the contortion definition, it can still be written
\begin{equation}
\Lw = \frac{h}{2 k^2} \left(\Kw^{\mu \nu \rho} \Kw_{\rho \nu \mu} -
\Kw^{\mu \rho}{}_\mu \Kw^\nu{}_{\rho \nu} \right).
\end{equation}

On the other hand, from Eq.~(\ref{relar}) it is possible to show that
\begin{equation}
- \Rbol = \Qw \equiv \left(\Kw^{\mu \nu \rho} \Kw_{\rho \nu \mu} -
\Kw^{\mu \rho}{}_\mu \Kw^\nu{}_{\rho \nu} \right) +
\partial_\mu (2 \, h \, \Tw^{\nu \mu}{}_\nu).
\end{equation}
Therefore, we see that
\begin{equation}
\Lw = \Lbol - \partial_\mu \left(2 \, h \, k^{-2} \,
\Tw^{\nu \mu}{}_\nu \right),
\end{equation}
where
\begin{equation}
\Lbol = - \frac{\sqrt{-g}}{2 k^2} \; \Rbol,
\end{equation}
represents the Einstein-Hilbert Lagrangian of general relativity, and where we have used the
identification $h = \sqrt{-g}$, with $g = \det(g_{\mu \nu})$. Up to a divergence, therefore,
the teleparallel Lagrangian is equivalent to the Einstein--Hilbert Lagrangian of general
relativity. It is interesting to observe that the first--order M{\o}ller's Lagrangian of
general relativity,\cite{moller}
\be
\Lbol_{\rm M} = \frac{h}{2 k^2} \left(\nabol_{\mu} h^{a \nu} \; \nabol_{\nu}
h_{a}{}^{\mu} - \nabol_{\mu} h_{a}{}^{\mu} \; \nabol_{\nu} h^{a \nu} \right),
\label{mlagra}
\ee
which differs from the Einstein-Hilbert Lagrangian by a total divergence, when rewritten in
terms of the Weitzeb\"ock connection coincides exactly---that is, without any boundary
term---with the teleparallel Lagrangian (\ref{gala}). Teleparallel gravity, therefore, can be
considered as fully equivalent with the M{\o}ller's first--order formulation of general
relativity.

Let us consider now the Lagrangian
\begin{equation}
{\mathcal L} = \Lw + {\mathcal L}_m
\end{equation}
where ${\mathcal L}_m$ is the Lagrangian of a general matter field $\Psi$. By performing
variations in relation to the gauge field $B^a{}_\rho$, we obtain the teleparallel version
of the gravitational field equation
\be
\partial_\sigma(h \Sw_a{}^{\rho \sigma}) -
k^2 \, (h \jw_{a}{}^{\rho}) = k^2 \, (h {\Theta}_{a}{}^{\rho}),
\label{tfe1}
\ee
where $\Sw_a{}^{\rho \sigma} = h_{a}{}^{\lambda} \Sw_{\lambda}{}^{\rho \sigma}$, and
\begin{equation}
h {\Theta}_{a}{}^{\rho} \equiv - \frac{\delta {\mathcal L}_m}{\delta
B^a{}_{\rho}} \equiv - \frac{\delta {\mathcal L}_m}{\delta h^a{}_{\rho}} = -
\left( \frac{\partial {\mathcal L}_m}{\partial h^a{}_{\rho}} -
\partial_\lambda \frac{\partial {\mathcal L}_m}{\partial_\lambda\partial h^a{}_{\rho}} \right)
\end{equation}
is the matter energy-momentum tensor. Analogously to the Yang-Mills theories,
\be
h \jw_{a}{}^{\rho} \equiv - \frac{\partial \Lw}{\partial B^a{}_{\rho}} \equiv -
\frac{\partial \Lw}{\partial h^a{}_{\rho}} =
\frac{h}{k^2} \, h_a{}^{\lambda} \, \Sw_c{}^{\nu \rho} \,
\Tw^c{}_{\nu \lambda} - h_a{}^{\rho} \Lw
\label{ptem1}
\ee
stands for the gauge current, which in this case represents the energy and momentum of the
gravitational field.\cite{gemt} In a purely spacetime form, it becomes
\be
\jw_\mu{}^{\rho} \equiv h^a{}_\mu \, \jw_{a}{}^{\rho} =
\frac{1}{k^2} \left( \Sw_\sigma{}^{\nu \rho} \,
\Tw^\sigma{}_{\nu \mu} - \frac{1}{4} \, \delta_\mu{}^\rho \,
\Sw_\sigma{}^{\nu \lambda} \, \Tw^\sigma{}_{\nu \lambda} \right),
\ee
which has the same structure of the symmetrized\cite{belinfante} energy--momentum tensor of
the electromagnetic field.\cite{landau}

Now, by using Eq.~(\ref{rela0}), the left-hand side of the field equation (\ref{tfe1}),
after a lengthy but straightforward calculation, can be shown to satisfy
\begin{equation}
\partial_\sigma(h \Sw_a{}^{\rho \sigma}) -
k^2 \, (h \jw_{a}{}^{\rho}) =
h \left({\stackrel{\circ}{R}}_a{}^{\rho} -
\frac{1}{2} \, h_a{}^{\rho} \;
{\stackrel{\circ}{R}} \right).
\end{equation}
As expected, due to the equivalence between the corresponding Lagrangians,
the teleparallel field equation (\ref{tfe1}) is equivalent to Einstein's equation
\be
{\stackrel{\circ}{R}}_a{}^{\rho} -
\frac{1}{2} \, h_a{}^{\rho} \,
{\stackrel{\circ}{R}} = k^2 \, {\Theta}_{a}{}^{\rho}.
\ee

\subsection{Gravitational Energy-Momentum Current}

The definition of an energy-momentum density for the gravitational field is one of the oldest
and most controversial problems of gravitation. As a true field, it would be natural to
expect that gravity should have its own local energy-momentum density. However, it is usually
asserted that such a density cannot be locally defined because of the equivalence
principle.\cite{mtw} As a consequence, any attempt to identify an energy-momentum density for
the gravitational field leads to complexes that are not true tensors. The first of such
attempt was made by Einstein who proposed an expression for the energy-momentum density of
the gravitational field which was nothing but the canonical expression obtained from
Noether's theorem.\cite{trautman} Indeed, this quantity is a pseudotensor, an object that
depends on the coordinate system. Several other attempts have been made, leading to different
expressions for the energy-momentum pseudotensor for the gravitational field.\cite{others}

Despite the existence of some controversial points related to the formulation of the
equivalence principle,\cite{synge} it seems true that, in the context of general relativity,
no tensorial expression for the gravitational energy-momentum density can exist. In spite of
some skepticism,\cite{mtw} there has been a continuous interest in this problem.\cite{recent}
In particular, a {\it quasilocal} approach\cite{broyork} has been proposed which is highly
clarifying.\cite{qlo} According to this approach, for each gravitational energy-momentum
pseudotensor, there is an associated {\it superpotential} which is a Hamiltonian boundary
term. The energy-momentum defined by such a pseudotensor does not really depend on the local
value of the reference frame, but only on the value of the reference frame on the boundary of
a region---then its {\it quasilocal} character. As the relevant boundary conditions are
physically acceptable, this approach validates the pseudotensor approach to the gravitational
energy-momentum problem.

However, in the gauge context of teleparallel gravity, the existence of a tensorial
expression for the gravitational energy-momentum density seems to be possible. Accordingly,
the absence of such expression should be attributed to the general relativity description of
gravitation, which seems not to be the appropriate framework to deal with this
problem.\cite{maluf95} In fact, as can be easily checked, the current $\jw_a{}^\rho$
transforms covariantly under a general spacetime coordinate transformation, is invariant
under local (gauge) translation of the tangent-space coordinates, and transforms covariantly
under a tangent--space Lorentz transformation. This means that $\jw_a{}^\rho$ is a true
spacetime and gauge tensor. Since our interest is the gravitational
energy-momentum current, let us consider the sourceless case, in which the gravitational
field equation becomes
\be
\partial_\sigma (h \Sw_a{}^{\rho \sigma}) -
k^2 \, (h \jw_{a}{}^{\rho}) = 0.
\label{tfe0}
\ee
Due to the anti-symmetry of $\Sw_a{}^{\rho \sigma}$ in the last two indices, $(h
\jw_{a}{}^{\rho})$ is conserved as a consequence of the field equation:
\be
\partial_\rho (h \jw_a{}^\rho) = 0.
\label{conser1}
\ee
Making use of the identity
\be
\partial_\rho h \equiv h \Gammaw^{\nu}{}_{\nu \rho} =
h \left( \Gammaw^{\nu}{}_{\rho \nu} - \Kw^{\nu}{}_{\rho \nu} \right),
\label{id1}
\ee
this conservation law can be rewritten in the manifestly covariant form
\be
\Dw_\rho \, \jw_a{}^\rho \equiv \partial_\rho \jw_a{}^\rho +
\left( \Gammaw^\rho{}_{\lambda \rho} - \Kw^\rho{}_{\lambda \rho} \right)
\jw_a{}^\lambda = 0,
\label{conser2}
\ee
with $\Dw_\rho$ the teleparallel covariant derivative (\ref{tcd}).

Let us find out now the relation between the current $\jw_a{}^\rho$ and the usual
gravitational energy-momentum pseudotensor. By using Eq.~(\ref{carco}) to express
$\partial_\rho h_a{}^\lambda$, the field equation (\ref{tfe0}) can be rewritten in
a purely spacetime form,
\be
\partial_\sigma(h \Sw_\lambda{}^{\rho \sigma}) -
k^2 \, (h \tw_{\lambda}{}^{\rho}) = 0,
\label{tfe2}
\ee
where
\be
h \tw_{\lambda}{}^{\rho} =
k^{-2} \, h \, \Gammaw^{\mu}{}_{\nu \lambda} \, \Sw_{\mu}{}^{\rho \nu}
- \delta_\lambda{}^{\rho} \Lw
\label{ptem2}
\ee
stands for the canonical energy-momentum pseudotensor of the gravitational
field.\cite{shirafuji96} It is important to notice that $\tw_{\lambda}{}^{\rho}$ is not
simply the gauge current $\jw_a{}^\rho$ with the algebraic index ``$a$'' changed to the
spacetime index ``$\lambda$''. It incorporates also an extra term coming from the
derivative term of Eq.~(\ref{tfe1}):
\be
\tw_\lambda{}^\rho = h^a{}_\lambda \, \jw_a{}^\rho +
k^{-2} \, \Gammaw^{\mu}{}_{\lambda \nu} \, \Sw_{\mu}{}^{\rho \nu}.
\label{ptem3}
\ee
We see thus clearly the origin of the connection-term which transforms the gauge current
$\jw_a{}^\rho$ into the energy-momentum pseudotensor $\tw_\lambda{}^\rho$. Through the same
mechanism, it is possible to appropriately exchange further terms between the derivative and
the current terms of the field equation (\ref{tfe2}), giving rise to different definitions
for the energy-momentum pseudotensor, each one connected to a different {\it superpotential}
$(h \Sw_\lambda{}^{\rho \sigma})$. It is important to remark finally that, like the gauge
current $(h \jw_a{}^\rho)$, the pseudotensor $(h \tw_\lambda{}^\rho)$ is conserved as a
consequence of the field equation:
\be
\partial_\rho (h \tw_\lambda{}^\rho) = 0.
\label{conser3}
\ee
However, in contrast to what occurs with $\jw_a{}^\rho$, due to the pseudotensor character
of $\tw_\lambda{}^\rho$, this conservation law cannot be rewritten in terms of the
teleparallel covariant derivative.

Because of its simplicity and transparency, the teleparallel approach to gravitation seems to
be much more appropriate than general relativity to deal with the energy problem of the
gravitational field. In fact, M{\o}ller already noticed a long time ago that a satisfactory
solution for the problem of the energy distribution in a gravitational field could be
obtained in the framework of a tetrad theory. In our notation, his expression for the
gravitational energy-momentum density is\cite{moller}
\be
h t_\lambda{}^\rho = \frac{\partial \Lbol_{\rm M}}{\partial \partial_\rho h^a{}_\mu} \;
\partial_\lambda h^a{}_\mu - \delta_\lambda{}^\rho \, \Lbol_{\rm M},
\label{moemt}
\ee
which is nothing but the usual Noether's canonical energy-mo\-men\-tum density in the tetrad
formulation of general relativity. Since M\o ller's Lagrangian, given by Eq.~(\ref{mlagra}),
is exactly (without any surface term) equivalent with the teleparallel Lagrangian
(\ref{gala}), the M{\o}ller's expression (\ref{moemt}) will correspond exactly with the
teleparallel energy-momentum density (\ref{ptem2}).

\subsection{Noether's Theorem: Matter Conservation Law}

Let us consider the action integral of a general matter field,
\be
{\mathcal S}_m = \frac{1}{c} \int {\mathcal L}_m \; d^4x.
\ee
We assume a first-order formalism, according to which the Lagrangian depends only on
the fields and on their first derivatives. Under a general transformation of the
spacetime coordinates,
\be
x^{\prime \rho} = x^\rho + \xi^\rho,
\label{stct}
\ee
the tetrad transforms according to
\be
\delta h_a{}^\mu = h_a{}^\rho \, \partial_\rho \xi^\mu -
\xi^\rho \, \partial_\rho h_a{}^\mu.
\label{htrans}
\ee
The corresponding transformation of the action integral is written as
\be
\delta {\mathcal S}_m = \frac{1}{c} \int {\Theta}_\mu{}^a \;
\delta h_a{}^\mu \; h \; d^4x,
\label{deltas2}
\ee
where $h \Theta_\mu{}^a = {\delta {\mathcal L}_m}/{\delta h_a{}^\mu}$ is the matter
energy-momentum tensor. Substituting $\delta h_a{}^\mu$ as given by Eq.~(\ref{htrans}), we
obtain
\be
\delta {\mathcal S}_m = \frac{1}{c} \int {\Theta}_\mu{}^a \, \left[
h_a{}^\rho \, \partial_\rho \xi^\mu -
\xi^\rho \, \partial_\rho h_a{}^\mu \right] h \; d^4x.
\label{ele}
\ee
Now, from the absolute parallelism condition (\ref{apc}), we have that
\be
\partial_\rho h_a{}^\mu = 0 -
\Gammaw^\mu{}_{\lambda \rho} \, h_a{}^\lambda.
\ee
Substituting into (\ref{ele}), after some manipulations, we get
\be
\delta {\mathcal S}_m = \frac{1}{c} \int \left[ {\Theta}_c{}^\rho \left(
\partial_\rho \xi^c + 0 \right) +
{\Theta}_\mu{}^\rho \Tw^\mu{}_{\lambda \rho} \xi^\lambda \right] h \; d^4x.
\ee
Contrary to some claims,\cite{hay72} provided the action ${\mathcal S}_m$ is local Lorentz
invariant, the energy--momentum tensor ${\Theta}_\mu{}^\rho$ is necessarily
symmetric.\cite{weinberg} Consequently, the above variation assumes the form
\be
\delta {\mathcal S}_m = \frac{1}{c} \int {\Theta}_c{}^\rho
\left[ \partial_\rho \xi^c + 0 -
\Kw^c{}_{b \rho} \xi^b \right] \; h \; d^4x.
\ee
Integrating the first term by parts, neglecting the surface term, and considering the
arbitrariness of $\xi^c$, it follows from the invariance of the action integral that
\be
\partial_\mu (h {\Theta}_a{}^\mu) - ( 0 - \Kw^{b}{}_{a \mu}) \;
h \, {\Theta}_b{}^\mu = 0.
\label{cola1}
\ee
Making use of the identity (\ref{id1}), the above conservation law becomes
\be
\partial_\mu {\Theta}_a{}^\mu +
(\Gammaw^\mu{}_{\rho \mu} - \Kw^\mu{}_{\rho \mu}) \; {\Theta}_a{}^\rho -
( 0 - \Kw^{b}{}_{a \mu}) \; {\Theta}_b{}^\mu = 0.
\label{cola2}
\ee
In a purely spacetime form, it becomes
\be
\partial_\mu \Theta_\lambda{}^\mu +
(\Gammaw^\mu{}_{\rho \mu} - \Kw^\mu{}_{\rho \mu}) \; {\Theta}_\lambda{}^\rho -
(\Gammaw^\rho{}_{\lambda \mu} - \Kw^\rho{}_{\lambda \mu}) \; {\Theta}_\rho{}^\mu
\equiv \dw_\mu {\Theta}_\lambda{}^\mu = 0.
\label{cola3}
\ee
This is the conservation law of matter energy-momentum tensor. In teleparallel gravity,
therefore, it is not the Weitzenb\"ock covariant derivative $\nablaw_\mu$, but the
teleparallel covariant derivative $\dw_\mu$ that yields the correct conservation law for the
energy-momentum tensors of matter fields. Of course, because of the relation (\ref{rela0}),
it can be written in the form
\be
{\stackrel{\circ}{\nabla}}{}_\mu \Theta_\lambda{}^\mu \equiv
\partial_\mu \Theta_\lambda{}^\mu +
{\stackrel{\circ}{\Gamma}}{}^\mu{}_{\rho \mu} \Theta_\lambda{}^\rho -
{\stackrel{\circ}{\Gamma}}{}^\rho{}_{\lambda \mu} \Theta_\rho{}^\mu = 0,
\label{grcon}
\ee
which is the corresponding conservation law of general relativity. It is important to remark
that these ``covariant conservation laws'' are not, strictly speaking, real conservation laws
in the sense that they do not yield a conserved ``charge''. They are actually identities,
called Noether identities, which govern the exchange of energy and momentum between the
matter and the gravitational fields.\cite{kopov} 

\subsection{Bianchi Identities}

Analogously to the Maxwell theory, the first Bianchi identity of the gauge theory for
the translation group is\cite{weitz}
\be
\partial_\rho \Tw^a{}_{\mu \nu} + \partial_\nu \Tw^a{}_{\rho \mu} +
\partial_\mu \Tw^a{}_{\nu \rho} = 0.
\label{bi1}
\ee
Through a tedious, but straightforward calculation, it can be rewritten in a purely
spacetime form:
\be
\Qw^{\rho}{}_{\theta \mu \nu} + \Qw^{\rho}{}_{\nu \theta \mu} +
\Qw^{\rho}{}_{\mu \nu \theta} = 0.
\label{ibq1}
\ee
Then, by making use of relation (\ref{relar}), it is easy to verify that it coincides
with the first Bianchi identity of general relativity:
\be
{\stackrel{\circ}{R}}{}^{\rho}{}_{\theta \mu \nu} +
{\stackrel{\circ}{R}}{}^{\rho}{}_{\nu \theta \mu} +
{\stackrel{\circ}{R}}{}^{\rho}{}_{\mu \nu \theta} = 0.
\label{birg1}
\ee

On the other hand, similarly to general relativity, teleparallel gravity presents also
a second Bianchi identity, which is given by
\be
\dw_\sigma \Qw_{\rho \theta \mu \nu} +
\dw_\nu \Qw_{\rho \theta \sigma \mu} +
\dw_\mu \Qw_{\rho \theta \nu \sigma} = 0.
\label{sbigr}
\ee
This identity is easily seen to be equivalent to the second Bianchi identity of general
relativity
\be
{\stackrel{\circ}{\nabla}}_\sigma {\stackrel{\circ}{R}}_{\rho \theta \mu \nu} +
{\stackrel{\circ}{\nabla}}_\nu {\stackrel{\circ}{R}}_{\rho \theta \sigma \mu} +
{\stackrel{\circ}{\nabla}}_\mu {\stackrel{\circ}{R}}_{\rho \theta \nu \sigma} = 0,
\label{birg2}
\ee
whose contracted form is
\be
{\stackrel{\circ}{\nabla}}_\rho \left[ {\stackrel{\circ}{R}}{}_\lambda{}^\rho -
\textstyle{\frac{1}{2}} \delta_\lambda{}^\rho {\stackrel{\circ}{R}} \right] = 0.
\ee
Through a similar procedure, the contracted form of the teleparallel Bianchi identity
(\ref{sbigr}) is found to be\cite{mg9}
\be
\dw_\rho \left[ \partial_\sigma (h \Sw_\lambda{}^{\rho \sigma}) -
k^2 \, (h \tw_{\lambda}{}^{\rho}) \right] = 0.
\label{bi3}
\ee
If we remember that, in the presence of a general source field, the teleparallel field
equation is given by
\be
\partial_\sigma(h \Sw_\lambda{}^{\rho \sigma}) -
k^2 \, (h \tw_{\lambda}{}^{\rho}) =
k^2 \, (h \Theta_{\lambda}{}^{\rho}),
\label{tfe3}
\ee
with ${\Theta}_{\lambda}{}^{\rho}$ the matter energy-momentum tensor, and taking into
account that
\be
\dw_\rho h = 0,
\ee
the Bianchi identity (\ref{bi3}) is seen to be consistent with the conservation law
\be
\dw_\rho {\Theta}_\lambda{}^\rho = 0,
\label{teleconbis}
\ee
as obtained from Noether's theorem [see Eq.~(\ref{cola3})].

\subsection{Role of Torsion in Teleparallel Gravity}

\subsubsection{Force Equation Versus Geodesics}

To begin with, let us consider, in the context of teleparallel gravity, the motion of a
spinless particle of mass $m$ in a gravitational field $B^{a}{}_{\mu}$. Analogously to
the electromagnetic case,\cite{landau} the action integral is written in the form
\be
{\mathcal S} = \int_{a}^{b} \left[ - m \, c \, d\sigma -
m \, c \, B^{a}{}_{\mu} \, u_{a} \, dx^{\mu} \right],
\label{acaop1}
\ee
where $d\sigma = (\eta_{a b} dx^a dx^b)^{1/2}$ is the Minkowski tangent-space invariant
interval,
\be
u^a = h^a{}_\mu \, u^\mu,
\ee
is the anholonomic particle four-velocity, with
\be
u^\mu = \frac{d x^\mu}{ds}
\label{ust}
\ee
the holonomic four-velocity, which is written in terms of the spacetime invariant interval
$ds$ = $(g_{\mu \nu} dx^\mu dx^\nu)^{1/2}$. It should be noticed that, in terms of
the tangent-space line element $d \sigma$, the four-velocity $u^a$ is holonomic:\cite{wep}
\be
u^a = \frac{d x^a}{d \sigma}.
\label{native}
\ee

The first term of the action (\ref{acaop1}) represents the action of a free particle, and the
second the coupling of the particle's mass with the gravitational field. Notice that the
separation of the action in these two terms is possible only in a gauge theory, like
teleparallel gravity, being not possible in general relativity. It is, however, equivalent
with the usual action of general relativity. In fact, if we introduce the identities
\be
h_a{}^\mu u^a u_\mu = 1
\ee
and
\be
\frac{\partial x^\mu}{\partial x^a} \, u^a \, u_\mu = \frac{d s}{d \sigma},
\ee
the action (\ref{acaop1}) can easily be seen to reduce to its general relativity version
\[
{\mathcal S} = - \int_{a}^{b} m \, c \, ds.
\]
In this case, the interaction of the particle with the gravitational field is
described by the metric tensor $g_{\mu \nu}$, which is present in $ds$.

Variation of the action (\ref{acaop1}) yields the equation of motion
\be
h^a{}_\mu \, \frac{d u_a}{d s} =
\Tw^a{}_{\mu \rho} \; u_a \, u^\rho.
\label{eqmot2}
\ee
This is the force equation governing the motion of the particle, in which the
teleparallel field strength $\Tw^a{}_{\mu \rho}$---that is, torsion---plays the role of
gravitational force. To write it in a purely spacetime form, we use the relation
\be
h^{a}{}_{\mu} \frac{d u_{a}}{d s} = \omega_\mu \equiv
\frac{d u_{\mu}}{d s} - \Gammaw^\theta{}_{\mu \nu} u_{\theta} \, u^{\nu},
\ee
where $\omega_\mu$ is the spacetime particle four--acceleration. We then get
\be
u^\nu \nablaw_\nu u_\mu \equiv \frac{d u_\mu}{d s} - \Gammaw^\theta{}_{\mu \nu} \;
u_\theta \; u^\nu = \Tw^\theta{}_{\mu \nu} \; u_\theta \; u^\nu.
\label{geode}
\ee
The left--hand side of this equation is the Weitzenb\"ock covariant derivative of
$u_\mu$ along the world-line of the particle. The presence of the torsion tensor
on its right--hand side, as already stressed, shows that in teleparallel gravity
torsion plays the role of gravitational force. By using the identity
\be
\Tw^\theta{}_{\mu \nu} \, u_\theta \, u^\nu = - \Kw^\theta{}_{\mu \nu}
\, u_\theta \, u^\nu,
\label{tuukuu}
\ee
this equation can be rewritten in the form
\be
u^\nu \dw_\nu u_\mu \equiv \frac{d u_\mu}{d s} - \left(\Gammaw^\theta{}_{\mu \nu} -
\Kw^\theta{}_{\mu \nu} \right) u_\theta \; u^\nu = 0.
\label{geode3}
\ee
The left--hand side of this equation is the teleparallel covariant derivative of
$u_\mu$ along the world-line of the particle. Using the relation (\ref{rela0}), it is
found to be
\be
u^\nu \nabol_\nu u_\mu \equiv \frac{d u_\mu}{d s} -
{\stackrel{\circ}{\Gamma}}{}^\theta{}_{\mu \nu} \; u_\theta \; u^\nu = 0.
\label{geo2}
\ee
This is precisely the geodesic equation of general relativity, which means that the
trajectories followed by spinless particles  are geodesics of the underlying Riemann
spacetime. In a locally inertial coordinate system, the first derivative of the metric tensor
vanishes, the Levi--Civita connection vanishes as well, and the geodesic equation
(\ref{geo2}) becomes the equation of motion of a free particle. This is the usual version of
the (strong) equivalence principle as formulated in general relativity.\cite{weinberg}

It is important to notice that, by using the torsion definition (\ref{tor}), the force
equation (\ref{geode}) can be written in the form
\be
\frac{d u_\mu}{d s} - \Gammaw^\theta{}_{\mu \nu} \; u_\theta \; u^\nu = 0.
\label{geodeflat}
\ee
As $\Gammaw_{\theta \nu \mu}$ is not symmetric in the last two indices, this is not a
geodesic equation. This means that the trajectories followed by spinless particles are not
geodesics of the induced Weitzenb\"ock spacetime. In a locally inertial coordinate system,
the first derivative of the metric tensor vanishes, and the Weitzenb\"ock connection
$\Gammaw_{\theta \nu \mu}$ becomes skew--symmetric in the first two indices. In this
coordinate system, therefore, owing to the symmetry of  $u^\theta \; u^\nu$, the force
equation (\ref{geodeflat}) becomes the equation of motion of a free particle. This is the
teleparallel version of the (strong) equivalence principle.\cite{paper1}

\subsubsection{Teleparallel Equivalent of the Kerr--Newman Solution}

In spite of the equivalence of teleparallel gravity with general relativity, there are
conceptual differences between these two theories. For example, whereas in general relativity
gravitation is described in terms of the curvature tensor, in teleparallel gravity it is
described in terms of torsion. In addition to the conceptual differences, there are also some
formal differences. For example, the possibility of decomposing torsion into three
irreducible parts under the group of global Lorentz transformations allows a better
understanding of the physical meaning of torsion. To exemplify this fact, we are going to
study in this section the teleparallel version of the Kerr-Newman solution. This solution has
a great generality as it reduces to the Kerr and to Schwarzschild solutions for some specific
values of its parameters.

We begin by computing the the Kerr-Newman tetrad in the Boyer-Lindquist coordinates, in
which the Kerr-Newman metric is written as\cite{mtw}
\be
d s^2 = g_{00} dt^2 + g_{11} dr^2 + g_{22} d\theta^2 +
g_{33} d\phi^2 + 2 g_{03} d\phi\; dt ,
\ee
where
\be
g_{00} = 1 - { R r \over \rho^2}, \;\;g_{11} = - { \rho^2 \over \Delta},\;\;
g_{22} = - \rho^2,
\ee
\be
g_{33} =  - \left( r^2 + a^2 + {R r a^2 \over  \rho^2} \sin^2{\theta} \right)
\sin^2{\theta},
\ee
\be
g_{03} = g_{30} =  {R r a \over \rho^2 }\sin^2{\theta}.
\ee
In these expressions, $\Delta = r^2 - R r + a^2$, $R=2m-q^2/r$, $\rho^2 = r^2 + a^2
\cos^2\theta$, with $a$ the angular momentum of a gravitational unit mass source, $m$
the mass of the solution, and $q$ the electric charge. For $q=0$ the Kerr--Newman metric
reduces to the Kerr metric, and for $a=q=0$ it reduces to the standard form of
the Schwarzschild solution.

Using the relation (\ref{gmn0}), it is possible to find the Kerr--Newman tetrad components.
They are
\be
h^{a}{}_{\mu} \equiv \pmatrix{
\gamma_{00} & 0 & 0 &\eta \cr
0 &\gamma_{11} \, {\rm s}\theta \, {\rm c}\phi &\gamma_{22}\,{\rm c}\theta \, {\rm c}\phi &
-\beta \, {\rm s}\phi \cr
0 &\gamma_{11} \, {\rm s}\theta \, {\rm s}\phi &\gamma_{22}\,{\rm c}\theta \, {\rm s}\phi
& \beta \, {\rm c}\phi \cr
0 &\gamma_{11} \,{\rm c}\theta  & - \gamma_{22}\,{\rm s}\theta & 0},
\label{tek1}
\ee
with the inverse tetrad given by
\be
h_{a}{}^{\mu} \equiv \pmatrix{
\gamma_{00}^{-1} & 0 & 0 & 0 \cr
-\beta \, g^{03} \, {\rm s}\phi &\gamma_{11}^{-1} \, {\rm s}\theta \, {\rm c}\phi
&\gamma_{22}^{-1}\,{\rm c}\theta \, {\rm c}\phi  & -\beta^{-1} \, {\rm s}\phi \cr
\beta \, g^{03} \, {\rm c}\phi&\gamma_{11}^{-1} \, {\rm s}\theta \, {\rm s}\phi
&\gamma_{22}^{-1}\, {\rm c}\theta \, {\rm s}\phi &\beta^{-1} \, {\rm c} \phi \cr
0 &\gamma_{11}^{-1} \, {\rm c}\theta &- \gamma_{22}^{-1}\,{\rm s}\theta & 0},
\label{tek2}
\ee
where $\beta^2 = \eta^{2} - g_{33} $, $\eta = g_{03}/\gamma_{00}$, and the notations
$\gamma_{00}=\sqrt{g_{00}}$, $\gamma_{ii}=\sqrt{- g_{ii}}$, $s\theta= \sin\theta$ and
$c\theta=\cos\theta$ have been introduced. Using Eqs.~(\ref{carco}) and
(\ref{tor}), through a lengthy, but straightforward
calculation, we obtain the non-zero components of the torsion tensor,
\ba
\Tw^{0}{}_{01} &=& -[\ln \sqrt{g_{00}}]_{,r} \nn \\
\Tw^{0}{}_{13} &=& \eta_{,r} /\gamma_{00} - kg^{03}(k_{,r} -
    \gamma_{11} \, {\rm s}\theta) \nn \\
\Tw^{0}{}_{23} &=& \eta_{,\theta}/\gamma_{00}-
    kg^{03}(k_{,\theta} - \gamma_{22} \, {\rm c}\theta) \nn \\
\Tw^{1}{}_{12} &=& -[\ln \sqrt{-g_{11}}]_{,\theta} \nn \\
\Tw^{2}{}_{12} &=& [\ln \sqrt{-g_{22}}]_{,r} - \gamma_{11}/\gamma_{22} \nn \\
\Tw^{3}{}_{13} &=& (k_{,r} - \gamma_{11} \, {\rm s}\theta)/k \nn \\
\Tw^{3}{}_{23} &=& ( k_{,\theta} - \gamma_{22} \, {\rm c}\theta)/k, \nn
\ea
where we have denoted ordinary derivatives by a ``comma''.

Now, as already mentioned, the torsion tensor can be decomposed in the form
\be
T_{\lambda \mu \nu} = \frac{2}{3} \, \left(t_{\lambda \mu \nu} -
t_{\lambda \nu \mu} \right) + \frac{1}{3} \, \left(g_{\lambda \mu} V_\nu -
g_{\lambda \nu} V_\mu \right) + \epsilon_{\lambda \mu \nu \rho} \, A^\rho,
\label{deco}
\ee
where $t_{\lambda \mu \nu}$ is the purely tensor part, and $V_\mu$ and $A^\rho$
represent respectively the vector and axial parts of torsion. They are defined by
\be
t_{\lambda \mu \nu} = \frac{1}{2} \, \left(T_{\lambda \mu \nu} +
T_{\mu\lambda \nu} \right) + \frac{1}{6} \, \left(g_{\nu \lambda} V_\mu +
g_{\nu \mu} V_\lambda \right) - \frac{1}{3} \, g_{\lambda \mu} \, V_\nu,
\label{pt1}
\ee
\be
V_{\mu} =  T^{\nu}{}_{\nu \mu},
\label{pt2}
\ee
\be
A^{\mu} = {1\over 6}\epsilon^{\mu\nu\rho\sigma}T_{\nu\rho\sigma}.
\label{pt3}
\ee
For the specific case of the Kerr--Newman solution, the non-zero components of the vector
torsion are
\[
\Vw_{1} = -[\ln \sqrt{g_{00}}]_{,r} - [\ln \sqrt{-g_{22}}]_{,r}
+ \gamma_{11}/\gamma_{22} - [\ln \beta]_{,r} + \gamma_{11} \, {\rm s}\theta/\beta,
\]
\[
\Vw_{2} = - [\ln \sqrt{-g_{11}}]_{,\theta} - [\ln \beta]_{,\theta} +
\gamma_{22} \, c\theta/\beta,
\]
whereas the non-zero components of the axial torsion are
\[
\Aw^{(1)}\times(6h) = -2 (g_{00}T^{0}{}_{23} + g_{03}T^{3}{}_{23})
\]
\[
\Aw^{(2)}\times(6h) = 2[g_{00}T^{0}{}_{13} + g_{03}(T^{3}{}_{13} +
T^{0}{}_{01})].
\]

To show the simplicity and transparency of teleparallel gravity, let us obtain the
equation of motion of the particle's spin vector. By expanding the metric components
up to first order in the angular momentum $a$, and by taking the weak-field limit,
characterized by keeping terms up to first order in $\alpha_m = 2m/r$ and
$\alpha_q=q^2/r^2$, the axial-vector tensor reduces to
\be
\Aw^{(1)}\times(6h) = - 2 (g_{03})_{,\theta}
\ee
\be
\Aw^{(2)}\times(6h) = 2 [\gamma_{00} \, (\eta)_{,r} - \eta \, (\gamma_{00})_{,r}],
\ee
where $h=r^2 \sin\theta$. Substituting the metric components, and keeping the weak-field
approximation, the space components $\mbox{{\boldmath $\Aw$}} = \Aw^{(1)} \gamma_{11} \,
{\bf e}_{r} + \Aw^{(2)} \gamma_{22} \, {\bf e}_{\theta}$ of the axial--vector torsion
becomes
\be
\mbox{{\boldmath $\Aw$}} \equiv \mbox{{\boldmath $\Aw_{m}$}}+\mbox{{\boldmath $\Aw_{q}$}} =
{\alpha_m \, a\over 3 r^{2}} [2 \cos{\theta} \; {\bf e}_{r} + \sin{\theta} \; {\bf
e}_{\theta} ] - {\alpha_q \, a\over 3 r^{2}} [2 \cos{\theta} \, {\bf e}_{r}
+ 2 \sin{\theta} \, {\bf e}_{\theta}],
\label{av1}
\ee
where $\mbox{{\boldmath $\Aw_{m}$}}$ and $\mbox{{\boldmath $\Aw_{q}$}}$ represent
respectively the mass and charge parts of the axial-vector tensor.

Let us consider now a Dirac particle in the presence of the axial--vector torsion
{\boldmath $\Aw$}. It has been shown by many authors\cite{hayshi,nit80,aud81} that the
particle spin ${\bf s}$ satisfies the equation of motion
\be
\frac{d{\bf s}}{dt} = - {\bf b} \times {\bf s},
\ee
where ${\bf b} = 3 \mbox{{\boldmath $A$}}/2$.
Using Eq.~(\ref{av1}), we get
\be
{\bf b} = {J \over  r^{3}} [2 \cos{\theta} \;
{\bf e}_{r} + \sin{\theta} \; {\bf e}_{\theta}]- \frac{(q^2/m) J}{r^{4}} \;
[2 \cos{\theta} \; {\bf e}_{r} + 2 \sin{\theta} \; {\bf e}_{\theta}],
\ee
with $J=ma$ the angular momentum. In the particular case of the Kerr so\-lu\-tion,
$q=0$, and one can write
\be
{\bf b} = {G \over r^{3}} \left[- \mbox{{\boldmath $J$}} +
3 (\mbox{{\boldmath $J$}} \cdot {\bf e}_{r}) \, {\bf e}_{r} \right],
\ee
where $\mbox{{\boldmath $J$}} = J {\bf e}_{z}$. This means that
\be
{\bf b} = {\bf \omega}_{LT},
\label{av2}
\ee
where ${\bf \omega}_{LT}$ is the Lense--Thirring precession angular velocity, which in
general relativity, as is well known, is produced by the gravitomagnetic component of the
gravitational field.\cite{cw} We see in this way that, in teleparallel gravity, the
axial--vector torsion $\mbox{\boldmath$A$}$ represents the gravitomagnetic component of the
gravitational field.\cite{teof} When $q\neq 0$, the electric charge contributes with an
additional term to the Lense--Thirring precession angular velocity.

\subsubsection{Dealing Without the Equivalence Principle}

As is well known, the electromagnetic interaction is not universal: there is no an
electromagnetic equivalence principle. In spite of this, Maxwell's theory, a gauge theory for
the unitary group $U(1)$, is able to consistently describe the electromagnetic interaction.
Given the analogy between electromagnetism and teleparallel gravity, the question then arises
whether the gauge approach of teleparallel gravity would also be able to describe the
gravitational interaction in the lack of universality, that is, in the absence of the weak
equivalence principle.

Let us then consider again the problem of the motion of a spinless particle in a
gravitational field represented by the translational gauge potential $B^{a}{}_{\mu}$,
supposing however that the gravitational mass $m_g$ and the inertial mass $m_i$ do not
coincide. In this case, the action integral is written in the form
\be
{\mathcal S} = \int_{a}^{b} \left( - m_i \, c \, d\sigma -
m_g \, c \, B^{a}{}_{\mu} \, u_{a} \, dx^{\mu} \right).
\label{acao1b}
\ee
Variation of the action (\ref{acao1b}) yields the equation of motion
\be
\left( \partial_\mu x^a +
\frac{m_g}{m_i} \; B^a{}_\mu \right) \frac{d u_a}{d s} =
\frac{m_g}{m_i} \; \Tw^a{}_{\mu \rho} \; u_a \, u^\rho.
\label{eqmot22}
\ee
This is the force equation governing the motion of the particle, in which the
teleparallel field strength $\Tw^a{}_{\mu \rho}$ plays the role of gravitational force.
Similarly to the electromagnetic Lorentz force, which depends on the relation
$q/m_i$, with $q$ the electric charge of the particle, the gravitational force
depends explicitly on the relation ${m_g}/{m_i}$ of the particle.

The crucial point is to observe that, although the equation of motion depends explicitly on
the relation $m_i/m_g$ of the particle, neither $B^a{}_\mu$ nor $\Tw^a{}_{\rho \mu}$ depends
on this relation. This means essentially that the teleparallel field equation (\ref{tfe1})
can be consistently solved for the gravitational potential $B^a{}_\mu$, which can then be
used to write down the equation of motion (\ref{eqmot22}), independently of the validity or
not of the weak equivalence principle. This means essentially that, even in the absence of
the weak equivalence principle, teleparallel gravity is able to describe the motion of a
particle with $m_g \neq m_i$.\cite{wep}

Let us now see what happens in the context of general relativity. By using the identity
(\ref{tuukuu}), the force equation (\ref{eqmot22}) can be rewritten in the form
\be
\frac{d u_\mu}{ds} - \Gammabol^\lambda{}_{\mu \rho} \, u_\lambda \, u^\rho =
\left(\frac{m_g - m_i}{m_g} \right) \partial_\mu x^a \, \frac{d u_a}{d s},
\label{eqmot6}
\ee
where use has been made also of the relation (\ref{rela0}). Notice that the violation of
the weak equivalence principle produces a deviation from the geodesic motion, which is
proportional to the difference between the gravitational and inertial masses. Notice
furthermore that, due to the assumed non-universality of free fall, there is no a local
coordinate system in which the gravitational effects are absent. Of course, when $m_g =
m_i$, the equation of motion (\ref{eqmot6}) reduces to the geodesic equation of general
relativity. However, in the absence of the weak equivalence principle, it is not a
geodesic equation, which means that it does not comply with the geometric description of
general relativity, according to which all trajectories must be given by geodesic
equations.

In order to reduce the force equation (\ref{eqmot22}) to a geodesic equation, it is
necessary to define a new tetrad field, which is given by
\be
\bar{h}^a{}_\mu = \partial_\mu x^a +
\frac{m_g}{m_i} \; B^a{}_\mu.
\ee
However, in this case the solution of the gravitational field equations, which in
general relativity is a tetrad or a metric tensor, would depend on the relation
${m_g}/{m_i}$ of the test particle, and this renders the theory inconsistent. We can
then conclude that, in the absence of the weak equivalence principle, the geometric
description of gravitation provided by general relativity breaks down. Since the gauge
potential $B^a{}_\mu$ can always be obtained independently of any property of the test
particle, teleparallel gravity remains as a consistent theory in the lack of universality.
Accordingly, $B^a{}_\mu$ can be considered as the most fundamental field representing
gravitation.\cite{wep}

\subsection{Phase Factor Approach to Teleparallel Gravity}

As we have just discussed, the fundamental field of teleparallel gravity is the gauge
potential $B^a{}_\mu$. In this formulation, gravitation becomes quite analogous to
electromagnetism. Based on this analogy, and relying on the phase-factor approach to
Maxwell's theory, a teleparallel nonintegrable phase-factor approach to gravitation can be
developed,\cite{global} which represents the quantum mechanical version of the classical
gravitational Lorentz force.

As is well known, in addition to the usual {\em differential} formalism, electromagnetism
presents also a {\em global} formulation in terms of a nonintegrable phase
factor.\cite{wy} According to this approach, electromagnetism can be considered as the
gauge invariant action of a nonintegrable (path-dependent) phase factor. For a particle
with electric charge $q$ traveling from an initial point ${\sf P}$ to a final point ${\sf
Q}$, the phase factor is given by
\be
\Phi_e({\sf P}|{\sf Q}) = \exp \left[\frac{i q}{\hbar c} \int_{\sf P}^{\sf Q}
A_\mu \, dx^\mu \right],
\ee
where $A_\mu$ is the electromagnetic gauge potential. In the classical
(non-quantum) limit, the nonintegrable phase factor approach yields the same results as
those obtained from the Lorentz force equation
\be
\frac{d u^a}{ds} = \frac{q}{m c^2} \, \Tw^a{}_b \, u^b.
\ee
In this sense, the phase-factor approach can be considered as the {\em quantum}
generalization of the {\em classical} Lorentz force equation. It is actually
more general, as it can be used both on simply-connected and on
multiply-connected domains. Its use is mandatory, for example, to describe the
Bohm-Aharonov effect, a quantum phenomenon taking place in a multiply-connected
space.

Now, in analogy with electromagnetism, $B^a{}_\mu$ can be used to construct a global
formulation for gravitation. To start with, let us notice that the electromagnetic
phase factor
$\Phi_e({\sf P}|{\sf Q})$ is of the form
\be
\Phi_e({\sf P}|{\sf Q}) = \exp \left[\frac{i}{\hbar} \, {\mathcal S}_e \right],
\ee
where ${\mathcal S}_e$ is the action integral describing the interaction of the charged
particle with the electromagnetic field. Now, in teleparallel gravity, the
action integral describing the interaction of a particle of mass $m$ with
gravitation, according to Eq.~(\ref{acaop1}), is given by
\be
{\mathcal S}_g = \int_{\sf P}^{\sf Q} m \, c \, B^a{}_\mu \, u_a \, dx^\mu.
\ee
Therefore, the corresponding gravitational nonintegrable phase factor turns out
to be\cite{global}
\be
\Phi_g({\sf P}|{\sf Q}) = \exp \left[\frac{i m c}{\hbar} \int_{\sf P}^{\sf Q}
B^a{}_\mu \, u_a \, dx^\mu \right].
\label{npf}
\ee
Similarly to the electromagnetic phase factor, it represents the {\em quantum}
mechanical law that replaces the {\em classical} gravitational Lorentz force equation
(\ref{eqmot2}). In fact, in the classical limit it yields the same results as the
gravitational Lorentz force.

The above global formulation for gravitation has already been applied to study the so
called Colella-Overhauser-Werner (COW) experiment,\cite{cow} yielding the correct quantum
phase-shift induced on the neutrons by their interaction with Earth's gravitational field.
It has also been applied to study the quantum phase-shift produced by the coupling of the
particle's kinetic energy with the gravitomagnetic component of the gravitational field,
which corresponds to the gravitational analog of the Aharonov-Bohm effect.\cite{gabere}

\section{Torsion Physics: Infinitely Many Equivalent Theories}
\label{gauge}

\subsection{Physical Motivation}

As stated in the Introduction, the classical equivalence between teleparallel gravity and
general relativity implies that curvature and torsion might be simply alternative ways of
describing the gravitational field, and consequently related to the same degrees of freedom
of gravity. Whether this interpretation for torsion is universal or not is an open question.
In other words, whether torsion has the same physical role in more general gravitation
theories, in which curvature and torsion are simultaneously present, is a question yet to be
solved. This is the problem we tackle next, where we review some attempts\cite{newt1,newt2}
to get a consistent answer to this puzzle.

Our approach will consist in studying the gravitational coupling prescription in the
presence of curvature and torsion, independently of the theory governing the dynamics of
the gravitational field. One has only to consider a gravitational field presenting
curvature and torsion, and use it to obtain, from an independent variational principle,
the particle or field equations in the presence of gravity.  This, however, is not an easy
task. The basic difficulty is that, differently from all other interactions of nature,
where covariance does determine the gauge connection, in the presence of curvature and
torsion, covariance {\em alone} is not able to determine the form of the gravitational
coupling prescription. The reason for this indefiniteness is that the space of Lorentz
connections is an affine space,\cite{koba} and consequently one can always add a tensor to
a given connection without destroying the covariance of the theory. As a result of this
indefiniteness, there will exist infinitely many possibilities for the gravitational
coupling prescription. Notice that in the specific cases of general relativity and
teleparallel gravity, characterized respectively by a vanishing torsion and a vanishing
curvature, the above indefiniteness is absent since in these cases the connections are
uniquely determined---and the corresponding coupling prescriptions completely
specified---by the requirement of covariance. Notice furthermore that in the case of
internal (Yang-Mills) gauge theories, where the concept of torsion is absent,\cite{foot3} the
above indefiniteness is not present either.

In order to consider the above problem, a strategy based on the equivalence principle will be
used. Notice that, due to the intrinsic relation of gravitation with spacetime, there is a
deep relationship between covariance (either under general coordinate or local Lorentz
transformations) and the equivalence principle. In fact, an alternative version of this
principle is the so called {\em principle of general covariance}.\cite{weinberg} It states
that a special relativity equation will hold in the presence of gravitation if it is
generally covariant, that is, it preserves its form under a general transformation of the
spacetime coordinates. Now, in order to make an equation generally covariant, a connection is
always necessary, which is in principle concerned only with the {\em inertial} properties of
the coordinate system under consideration. Then, using the equivalence between inertial and
gravitational effects, instead of representing inertial properties, this connection can
equivalently be assumed to represent a {\em true gravitational field}. In this way, equations
valid in the presence of gravitation are obtained from the corresponding special relativity
equations. Of course, in a locally inertial coordinate system, they must go back to the
corresponding equations of special relativity. The principle of general covariance,
therefore, can be considered as an {\em active} version of the equivalence principle in the
sense that, by making a special relativity equation covariant, and by using the strong
equivalence principle, it is possible to obtain its form in the presence of gravitation. It
should be emphasized that the general covariance alone is empty of any physical content as
any equation can be made covariant. Only when use is made of the strong equivalence
principle, and the inertial compensating term is assumed as representing a true gravitational
field, principle of general covariance can be seen as a version of the equivalence
principle.\cite{sciama}

The above description of the general covariance principle refers to its usual {\it holonomic}
version. An alternative, more general version of the principle can be obtained in the context
of nonholonomic frames. The basic difference  between these two versions is that, instead of
requiring that an equation be covariant under a general transformation of the spacetime
coordinates, in the nonholonomic-frame version the equation is required to transform
covariantly under a {\em local} Lorentz rotation of the frame. Of course, in spite of the
different nature of the involved transformations, the physical content of both approaches are
the same.\cite{lorentz} The frame version, however, is more general in the sense that,
contrary to the coordinate version, it holds for integer as well as for half-integer spin
fields.

The crucial point now is to observe that, when the {\em purely inertial} connection is
replaced by a connection representing a {\em true gravitational field}, the principle of
general covariance naturally defines a covariant derivative, and consequently also a
gravitational coupling prescription. For the cases of general relativity and teleparallel
gravity, the nonholonomic-frame version of this principle has already been seen to yield the
usual coupling prescriptions of these theories.\cite{mospe} Our purpose here will then be to
determine, in the general case characterized by the simultaneous presence of curvature and
torsion, the form of the gravitational coupling prescription {\em implied by the general
covariance principle}.

\subsection{Nonholonomic General Covariance Principle}

Let us consider the Minkowski spacetime of special relativity, endowed with the Lorentzian
metric $\eta$. In this spacetime one can take the frame
$\delta_a=\delta_a{}^\mu\partial_\mu$ as being a trivial (holonomous) tetrad, with
components $\delta_a{}^\mu$. Consider now a {\em local}, that is, point-dependent Lorentz
transformation $\Lambda_a{}^b=\Lambda_a{}^b(x)$. It yields the new frame
\begin{equation}
h_{a} = h_{a}{}^{\mu} \partial_\mu,
\label{e_a}
\end{equation}
with components $h_a{}^\mu \equiv h_a{}^\mu(x)$ given by
\begin{equation}
h_{a}{}^{\mu} = \Lambda_{a}{}^{b} \; \delta_{b}{}^{\mu }.
\label{deltabar}
\end{equation}
Notice that, on account of the locality of the Lorentz transformation,
the new frame $h_a$ is nonholonomous, with $f^a{}_{bc}$ as coefficient of nonholonomy [see
Eq.~(\ref{tenm})]. So, if one makes use of the orthogonality property of the tetrads, we
see from Eq.~(\ref{deltabar}) that the Lorentz group element can be written in the form
$\Lambda_b{}^d = h_b{}^\rho \delta_\rho{}^d$. From this expression, it follows that
\begin{equation}
(h_a \Lambda_b{}^d)\Lambda^c{}_d =
\frac{1}{2} \left(f_b{}^c{}_a + f_a{}^c{}_b - f^c{}_{ba} \right).
\label{deltabar e f}
\end{equation}

On the other hand, the action describing a free particle in Minkowski spacetime is
[see Eq.~(\ref{acaop1})]
\begin{equation}
S = - m c \int d\sigma.
\label{action}
\end{equation}
Seen from the {\it holo\-no\-mous} frame $\delta_a$, the corresponding equation of motion
is
\begin{equation}
\frac{d v^a}{d s} = 0,
\label{plivre4}
\end{equation}
where $v^a = \delta^a{}_\mu u^\mu$, with $u^\mu = (dx^{\mu}/ds)$ the holonomous
particle four-velocity. Seen from the {\it nonholonomous} frame $h_a$, a straightforward
calculation shows that the equation of motion (\ref{plivre4}) is
\begin{equation}
\frac{du^{c}}{ds} + \frac{1}{2}
\left(f_b{}^c{}_a + f_a{}^c{}_b - f^c{}_{ba} \right) u^a u^b = 0,
\label{plivre5}
\end{equation}
where $u^c = \Lambda^c{}_d \, v^d = h^c{}_\mu u^\mu$, and use has been made of
Eq.~(\ref{deltabar e f}). It is important to emphasize that, in the
flat spacetime of special relativity one is free to choose any tetrad $\{e_{a}\}$ as
a moving frame. The fact that, for each $x \in M$, the frame $h_{a}$ can be
arbitrarily rotated introduces the compensating term $\frac{1}{2} (f_b{}^c{}_a +
f_a{}^c{}_b - f^c{}_{ba} )$ in the free-particle equation of motion. This term, therefore,
is concerned only with the inertial properties of the frame.

\subsubsection{Equivalence Between Inertial and Gravitational Effects}

According to the general covariance principle, the equation of motion valid in the
presence of gravitation can be obtained from the corresponding special relativistic
equation by replacing the inertial compensating term by a connection
$A^c{}_{ab}$ representing a true gravitational field. Considering a general Lorentz-valued
connection presenting both curvature and torsion, one can write\cite{abp1}
\begin{equation}
A^c{}_{ba} - A^c{}_{ab} = T^c{}_{ab} + f^c{}_{ab},
\end{equation}
with $T^c{}_{ba}$ the torsion of the connection $A^c{}_{ab}$. Use of this
equation for three different combination of indices gives
\begin{equation}
A^c{}_{ab} =
\frac{1}{2} \left( f_b{}^c{}_a + f_a{}^c{}_b - f^c{}_{ba} \right) +
\frac{1}{2} \left( T_b{}^c{}_a + T_a{}^c{}_b - T^c{}_{ba} \right).
\label{genecom}
\end{equation}
Accordingly, the compensating term of Eq.~(\ref{plivre5}) can be written in the form
\begin{equation}
\frac{1}{2} \left( f_b{}^c{}_a + f_a{}^c{}_b - f^c{}_{ba} \right) =
A^c{}_{ab} - K^c{}_{ab}.
\label{equiva}
\end{equation}
This equation is actually an expression of the
equivalence principle. In fact, whereas its left-hand side involves only {\it
inertial} properties of the frames, its right-hand side contains purely {\it
gravitational} quantities. Using this expression in Eq.~(\ref{plivre5}), one gets
\begin{equation}
\frac{du^{c}}{ds} + A^c{}_{ab} \, u^a \, u^b = K^c{}_{ab} \, u^a \, u^b.
\label{plivre5c}
\end{equation}
This is the particle equation of motion in the presence of curvature and torsion that
follows from the principle of general covariance. It entails a very peculiar
interpretation for contortion, which appears playing the role of a gravitational
force.\cite{paper1} Because of the identity (\ref{rela00}), it is easy to see that the
equation of  motion (\ref{plivre5c}) is equivalent with the geodesic equation of general
relativity:
\begin{equation}
\frac{du^{c}}{ds} + \Abol^c{}_{ab} \, u^a \, u^b = 0.
\label{plivre6}
\end{equation}

\subsection{Gravitational Coupling Prescription}

The equation of motion (\ref{plivre5c}) can be written in the form
\begin{equation}
u^\mu \D_\mu u^c = 0,
\label{cova2}
\end{equation}
with $\D_\mu$ a generalized Fock--Ivanenko derivative, which when acting on a general
vector field $X^{c}$ reads
\begin{equation}
\D_\mu X^c =
\partial_\mu X^c + (A^c{}_{a\mu} - K^c{}_{a\mu}) \; X^a.
\label{gfic}
\end{equation}
We notice in passing that this covariant derivative satisfies the relation
\begin{equation}
\D_\mu X^c = h^c{}_\rho \, D_\mu X^\rho,
\end{equation}
where
\begin{equation}
D_\mu X^\rho = \partial_\mu X^\rho + (\Gamma^\rho{}_{\lambda \mu} -
K^\rho{}_{\lambda \mu}) \; X^\lambda
\label{stfi}
\end{equation}
is the corresponding spacetime derivative.

Now, using the vector representation (\ref{vecre}) of the Lorentz generators, the
generalized Fock--Ivanenko derivative (\ref{gfic}) can be written in the form
\begin{equation}
\D_{\mu} X^{c} =
\partial_{\mu} X^{c} - \frac{i}{2} (A^{ab}{}_{\mu} - K^{ab}{}_{\mu}) \;
(S_{ab})^c{}_d \; X^d.
\label{acomivetor2}
\end{equation}
Furthermore, although obtained in the case of a Lorentz vector field (four-velocity), the
compensating term (\ref{deltabar e f}) can be easily verified to be the same for any
field. In fact, denoting by $U \equiv U(\Lambda)$ the element of the Lorentz group in an
arbitrary representation, it can be shown that
\begin{equation}
(h_{a} U) U^{-1} = - \frac{i}{4}
\left(f_{bca} + f_{acb} - f_{cba} \right) \, J^{bc},
\end{equation}
with $J^{bc}$ denoting the corresponding Lorentz generator. In this case, the covariant
derivative (\ref{acomivetor2}) will have the form
\begin{equation}
\D_{\mu} = \partial_{\mu} - \frac{i}{2} \left(A^{ab}{}_{\mu} - K^{ab}{}_{\mu} \right)
J_{ab}.
\label{genecova}
\end{equation}
This means that, in the presence of curvature and torsion, the coupling
prescription of fields carrying an arbitrary representation of the Lorentz group
will be
\begin{equation}
\partial_a \equiv \delta^\mu{}_a \partial_\mu \rightarrow
\D_a \equiv e^\mu{}_a \D_\mu.
\end{equation}
Of course, due to the relation (\ref{rela00}), it is clearly equivalent with the
coupling prescription of general relativity.

\subsection{The Connection Space}

\subsubsection{Defining Translations}

The mathematical validity of the coupling prescription (\ref{genecova}) is rooted on the
fact that a general connection space is an infinite, homotopically trivial affine
space.\cite{singer} In the specific case of Lorentz connections, a point in this space will
be a connection
\begin{equation}
A = A^{b c} {}_\mu\, J_{b c} \; dx^\mu
\end{equation}
presenting simultaneously curvature and torsion. In the language of differential forms,
they are defined respectively by
\begin{equation}
R = dA + A A \equiv \D_A A
\label{fcurve}
\end{equation}
and
\begin{equation}
T = dh + A h \equiv \D_A h,
\label{ftorsion}
\end{equation}
where  $\D_A$ denotes the covariant differential in the connection $A$.
Now, given two connections $A$ and $\bar{A}$, the difference
\begin{equation}
k = \bar{A} - A
\label{trans1}
\end{equation}
is also a 1-form assuming values in the Lorentz Lie algebra, but transforming
covariantly:
\begin{equation}
k = U k U^{-1}.
\end{equation}
Its covariant derivative is consequently given by
\begin{equation}
\D_A k = dk + \{A, k \}.
\end{equation}
It is then easy to verify that, given two connections such that $\bar{A}=A+k$,
their curvature and torsion will be related by
\begin{equation}
\bar{R} = R + \D_A k + k \, k
\end{equation}
and
\begin{equation}
\bar{T} = T + k \, h.
\end{equation}
The effect of adding a covector $k$ to a given connection $A$, therefore, is to
change its curvature and torsion 2-forms.

We rewrite now Eq.~(\ref{trans1}) in components:
\begin{equation}
A^a{}_{bc} = \bar{A}^a{}_{bc} - k^a{}_{bc}.
\end{equation}
Since $k^a{}_{bc}$ is a Lorentz-valued covector, it is necessarily anti-symmetric in
the {\it first two} indices. Separating $k^a{}_{bc}$ in the symmetric and
anti-symmetric parts in the {\it last two} indices, one gets
\begin{equation}
k^a{}_{bc}  = \onehalf (k^a{}_{bc}  + k^a{}_{cb} ) +
\onehalf (k^a{}_{bc}  - k^a{}_{cb}).
\end{equation}
Defining
\begin{equation}
k^a{}_{bc}  - k^a{}_{cb} \equiv t^a{}_{cb} = - t^a{}_{bc},
\end{equation}
and using this equation for three combination of indices, it is easy to verify that
\begin{equation}
k^a{}_{bc} = \onehalf (t_a{}^c{}_b + t_b{}^c{}_a - t^a{}_{bc} ).
\label{contorafim}
\end{equation}
This means essentially that the difference between any two Lorentz-valued connections has
the form of a contortion tensor.

\subsubsection{Equivalence under Translations in the Connection Space}

As already discussed, due to the affine character of the connection space, one can always add
a tensor to a given connection without spoiling the covariance of the derivative
(\ref{genecova}). Since adding a tensor to a connection corresponds just to redefining the
origin of the connection space, this means that covariance does not determine a preferred
origin for this space. Let us then analyze the physical meaning of translations in the
connection space. To begin with, we take again the connection appearing in the covariant
derivative (\ref{genecova}):
\begin{equation}
\Omega^a{}_{bc}  \equiv A^a{}_{bc} - K^a{}_{bc}.
\label{equiva2}
\end{equation}
A translation in the
connection space with parameter $k^a{}_{bc}$ corresponds to
\begin{equation}
\bar{\Omega}^a{}_{bc} = \Omega^a{}_{bc} + k^a{}_{bc} \equiv
A^a{}_{bc} - K^a{}_{bc} + k^a{}_{bc}.
\label{equiva3}
\end{equation}
Now, since $k^a{}_{bc}$ has always the form of a contortion tensor, as
given by Eq.~(\ref{contorafim}), the above connection is equivalent to
\begin{equation}
\bar{\Omega}^a{}_{bc} = A^a{}_{bc} - \bar{K}^a{}_{bc},
\end{equation}
with $\bar{K}^a{}_{bc} = K^a{}_{bc} - k^a{}_{bc}$ another contortion tensor.

Let us then consider a few particular cases. First, we choose  $t^a{}_{bc}$ as the
torsion of the connection $A^a{}_{bc}$, that is, $t^a{}_{bc} = T^a{}_{bc}$. In
this case, $k^a{}_{bc} = K^a{}_{bc}$, and the last two terms of Eq.~(\ref{equiva3}) cancel
each other, yielding
$\bar{K}^a{}_{bc} = 0$. This means that the torsion of ${A}^a{}_{bc}$ vanishes,
and we are left with the torsionless spin connection of general relativity:
\begin{equation}
\bar{\Omega}^a{}_{bc} = \Abol^a{}_{bc}.
\label{einstein}
\end{equation}
On the other hand, if we choose $t^a{}_{bc}$ such that
\begin{equation}
t^a{}_{bc} = T^a{}_{bc} - f^a{}_{bc},
\end{equation}
the connection $A^a{}_{bc}$ vanishes, which characterizes tele\-parallel gravity.
In this case, the resulting connection has the form\cite{tsc}
\begin{equation}
\bar{A}^a{}_{bc} = - \Kw^a{}_{bc},
\label{weitzen}
\end{equation}
where $\Kw^a{}_{bc}$ is the contortion tensor written in terms of the Weitzenb\"ock torsion
$\Tw^a{}_{bc} = - f^a{}_{bc}$. There are actually infinitely many choices for
$t^a{}_{bc}$, each one defining a different translation in the connection space, and
consequently yielding a connection $A^a{}_{bc}$ with different curvature and torsion. All
cases, however, are ultimately equivalent with general relativity as for all cases the
{\it dynamical} spin connection is
\begin{equation}
\bar{\Omega}^a{}_{bc} = A^a{}_{bc} - \bar{K}^a{}_{bc} \equiv
\Abol^a{}_{bc}.
\end{equation}
It is important to emphasize that, despite yielding physically equivalent coupling
prescriptions, the physical equations are not covariant under a translation in the
connection space. For example, under a particular translation, the geodesic equation of
general relativity is lead to the force equation of teleparallel gravity, which are
completely different equations. These two equations, however, as well as any other
obtained through a general translation in the connection space, are equivalent in the
sense that they describe the same physical trajectory.

\subsection{The Spinning Particle}

As an application of the gravitational minimal coupling prescription entailed by the
covariant derivative (\ref{genecova}), let us study the motion of a classical particle of
mass $m$ and spin {\bf s} in a gravitational field presenting curvature and
torsion.\cite{newt2} According to the gauge approach, the action integral describing such a
particle {\it minimally} coupled to gravitation is
\begin{equation}
{\mathcal S} = \int_{a}^{b} \left[ - m \, c \, d\sigma -
\frac{1}{c^{2}} B^{a}{}_{\mu} \, {\mathcal P}_{a} \, dx^{\mu } +
\frac{1}{2} \, \Omega^{ab}{}_{\mu} \, {\mathcal S}_{ab} \, dx^{\mu}\right],
\label{acao3}
\end{equation}
where ${\mathcal P}_{a} = m c u_a$ is the Noether charge associated with the invariance
of ${\mathcal S}$ under translations,\cite{paper1} and ${\mathcal S}_{ab}$ is the Noether
charge associated with the invariance of ${\mathcal S}$ under Lorentz
transformations.\cite{drech80} In other words, ${\mathcal P}_{a}$ is the momentum, and
${\mathcal S}_{ab}$ is the spin angular momentum density, which satisfies the Poisson
relation
\begin{equation}
\left\{ {\mathcal S}_{ab}, {\mathcal S}_{cd}\right\} = \eta_{ac} \, {\mathcal
S}_{bd} + \eta_{bd} \, {\mathcal S}_{ac} - \eta_{ad} \, {\mathcal S}_{bc} -
\eta_{bc} \, {\mathcal S}_{ad}.
\label{poisson}
\end{equation}
Notice that, according to this prescription, the particle's momentum couples minimally to
the translational gauge potential $B^{a}{}_{\mu}$, whereas the spin of the particle couples
minimally to the dynamical spin connection $\Omega^{ab}{}_{\mu}$, which is nothing but the
Ricci coefficient of rotation:
\[
\Omega^{ab}{}_{\mu} = (A^{ab}{}_{\mu} - K^{ab}{}_{\mu})
\equiv \Abol^{ab}{}_{\mu}.
\]

The Routhian arising from the action (\ref{acao3}) is
\begin{equation}
{\mathcal R}_0 = - m \, c \, \sqrt{u^{2}} \ \frac{d\sigma}{ds} -
\frac{1}{c^{2}} \, B^{a}{}_{\mu} \, {\mathcal P}_{a} \, u^{\mu} +
\frac{1}{2} \Omega^{ab}{}_{\mu} \, {\mathcal S}_{ab} \, u^{\mu },
\label{r0}
\end{equation}
where the weak constraint $\sqrt{u^{2}} \equiv \sqrt{u_a u^a}$ = $\sqrt{u_\mu u^\mu}$ =
$1$ has been introduced in the first term. The equation of motion for the particle
trajectory is obtained from
\begin{equation}
\frac{\delta}{\delta x^{\mu}} \int {\mathcal R}_0 \, ds = 0,
\label{euler}
\end{equation}
whereas the equation of motion for the spin tensor follows from
\begin{equation}
\frac{d{\mathcal S}_{ab}}{ds} = \{{\mathcal R}_0, {\mathcal S}_{ab}\}.
\label{eqspin1}
\end{equation}

Now, the four-velocity and the spin angular momentum density must satisfy the
constraints
\begin{eqnarray}
{\mathcal S}_{ab} {\mathcal S}^{ab} &=& 2 {\bf s}^{2} \label{v2} \\
{\mathcal S}_{ab}u^{a} &=& 0.
\label{v3}
\end{eqnarray}
However, since the equations of motions obtained from the Routhian ${\mathcal R}_0$ do
not satisfy the above constraints, it is necessary to include them into the Routhian. The
simplest way to achieve this amounts to the following.\cite{by} First, a new expression for
the spin is introduced:
\begin{equation}
\tilde{{\mathcal S}}_{ab} = {\mathcal S}_{ab} -
\frac{{\mathcal S}_{ac} u^{c} u_{b}}{u^{2}} -
\frac{{\mathcal S}_{cb} u^{c} u_{a}}{u^{2}}.
\label{stil}
\end{equation}
This new tensor satisfies the Poisson relation (\ref{poisson}) with the
metric $\eta _{ab} - u_{a} u_{b}/u^{2}$. A new Routhian that incorporates the
above constraints is obtained by replacing all ${\mathcal S}_{ab}$ in ${\mathcal R}_0$ by
$\tilde{{\mathcal S}}_{ab}$, and by adding to it the term
\[
\frac{du^{a}}{ds} \frac{{\mathcal S}_{ab} u^{b}}{u^{2}}.
\]
The new Routhian is then
\begin{equation}
{\mathcal R} = - m \, c \, \sqrt{u^{2}} \, \frac{d\sigma}{ds} -
\frac{1}{c^{2}} B^{a}{}_{\mu} \, {\mathcal P}_{a} \, u^{\mu} +
\frac{1}{2} \Omega^{ab}{}_{\mu} \, {\mathcal S}_{ab} \, u^{\mu } -
\frac{{\mathcal D} u^{a}}{{\mathcal D}s} \frac{{\mathcal S}_{ab} u^{b}}{u^{2}},
\label{rou}
\end{equation}
where
\[
\frac{{\mathcal D} u^{a}}{{\mathcal D}s} = u^\mu \, {\mathcal D}_\mu u^a,
\]
with ${\mathcal D}_\mu$ the covariant derivative (\ref{genecova}).

Using the Routhian (\ref{rou}), the equation of motion for the spin is found to be
\be
\frac{{\mathcal D} {\mathcal S}_{ab}}{{\mathcal D}s} = \left( u_{a} \, {\mathcal S}_{bc} -
u_{b} \, {\mathcal S}_{ac} \right) \frac{{\mathcal D} u^{c}}{{\mathcal D}s},
\ee
which coincides with the corresponding result of general relativity. Making use of the
Lagrangian formalism, the next step is to obtain the equation of motion defining the
trajectory of the particle. Through a tedious but straightforward calculation, it is
found to be
\begin{equation}
\frac{\mathcal D}{{\mathcal D} s} \left( m \, c \, u_c \right) +
\frac{\mathcal D}{{\mathcal D} s} \left(
\frac{{\mathcal D} u^{a}}{{\mathcal D}s}\frac{{\mathcal S}_{ac}}{u^{2}} \right) =
\frac{1}{2} (R^{ab}{}_{\mu \nu} - Q^{ab}{}_{\mu \nu}) \,
{\mathcal S}_{ab} \, u^\nu \, h_c{}^\mu,
\label{eq1}
\end{equation}
where
\be
Q^a{}_{b \mu \nu} = \D_{\mu}{}K^a{}_{b \nu} -
\D_{\nu}{}K^a{}_{b \mu} +  K^a{}_{d \mu}
\; K^d{}_{b \nu} - K^a{}_{d \nu} \;
K^d{}_{b \mu}.
\label{qdkgeral}
\ee
Using the constraints (\ref{v2}-\ref{v3}), it is
easy to verify that
\[
\frac{{\mathcal D} u^{a}}{{\mathcal D}s}
\frac{{\mathcal S}_{ac}}{u^{2}} = u^a \,
\frac{\mathcal D {\mathcal S}_{c a}}{\mathcal D s}.
\]
As a consequence, Eq.~(\ref{eq1}) acquires the form
\begin{equation}
\frac{\mathcal D}{{\mathcal D} s}\left(m c u_c +
u^a \frac{{\mathcal D} {\mathcal S}_{c a}}{{\mathcal D} s} \right) =
\frac{1}{2} (R^{ab}{}_{\mu \nu} - Q^{ab}{}_{\mu \nu}) \,
{\mathcal S}_{ab} \, u^\nu \, h_c{}^\mu.
\label{eq1.5}
\end{equation}
Defining the generalized four-momentum
\be
{\mathbb P}_{c} = h_c{}^\mu \, {\mathbb P}_\mu \equiv
m \, c \, u_{c} + u^{a} \, \frac{{\mathcal D} {\mathcal S}_{c a}}{{\mathcal D}s},
\ee
we get
\begin{equation}
\frac{{\mathcal D} {\mathbb P}_\mu}{{\mathcal D} s} =
\frac{1}{2} \, (R^{ab}{}_{\mu \nu} - Q^{ab}{}_{\mu \nu}) \,
{\mathcal S}_{ab} \, u^\nu.
\label{eq2}
\end{equation}
This is the equation governing the motion of the particle in the presence of both
curvature and torsion. It is written in terms of a general spin connection, as well as
in terms of its curvature and torsion. It can be rewritten in terms of the Ricci
coefficient of rotation only, in which case it reduces to the ordinary Pa\-pa\-pe\-trou
equation\cite{papapetrou}
\begin{equation}
\frac{\Dbol {\mathbb P}_\mu}{{\mathcal D} s} =
\frac{1}{2} \, \Rbol^{ab}{}_{\mu \nu} \, {\mathcal S}_{ab} \, u^\nu.
\end{equation}
It can also be rewritten in terms of the
teleparallel spin connection (\ref{tsc}), in which case it reduces to the teleparallel
equivalent of the Papapetrou equation,
\begin{equation}
\frac{\Dw {\mathbb P}_\mu}{{\mathcal D} s} = -
\frac{1}{2} \, \Qw^{a b}{}_{\mu \nu} \, {\mathcal S}_{ab} \, u^\nu,
\label{tepapa}
\end{equation}
with $\Qw^{ab}{}_{c d}$ given by Eq.~(\ref{qdk}). Notice that the particle's spin, similarly
to the electromagnetic field [see the teleparallel Maxwell equation (\ref{amax2})], couples
to a curvature-like tensor, which is a tensor written in terms of torsion only. It is
important to recall that, although physically equivalent, there are conceptual differences in
the way the above equations of motion describe the gravitational interaction.

\section{Outlook and Perspectives}

This review is made up of two main parts. In the first, represented by Section~\ref{telegra},
a comprehensive description of the teleparallel equivalent of general relativity was
presented. In spite of the mentioned equivalence, there are conceptual differences between
general relativity and teleparallel gravity, the most significant being the different
character of the fundamental field of the theories: whereas in general relativity it is a
tetrad field $h^a{}_\mu$ (or equivalently, a metric tensor $g_{\mu \nu}$), in teleparallel
gravity it is a translational gauge potential $B^a{}_\mu$, the nontrivial part of the tetrad
field:
\be
h^a{}_\mu = \partial_\mu x^a + B^a{}_\mu.
\label{ft1}
\ee
This apparently small difference has deep consequences. In fact, any gravitational theory
whose fundamental field is a tetrad (or a metric), is necessarily a {\em geometrical theory}.
On the other hand, a theory whose fundamental field is a gauge potential, like for example
teleparallel gravity, is non--geometrical in essence. It is actually a gauge theory, and as
such it is able to describe the gravitational interaction in the absence of the weak
equivalence principle.

To understand this point, let us consider a particle whose gravitational mass $m_g$ does not
coincide with its inertial mass $m_i$. In this case, a geometrical theory for gravitation
would require the introduction of a new tetrad field, given by\cite{wep}
\be
\bar{h}^a{}_\mu = \partial_\mu x^a + \frac{m_g}{m_i} \, B^a{}_\mu.
\label{ft2}
\ee
Since the relation ${m_g}/{m_i}$ of the test particle appears ``inside'' the tetrad
definition, any theory in which $\bar{h}^a{}_\mu$ is the fundamental field will be
inconsistent in the sense that particles with different relations ${m_g}/{m_i}$ will require
different solutions of the field equations to keep a geometric description of gravitation, in
which the trajectories are necessarily given by geodesics. On the other hand, we see from the
tetrad (\ref{ft2}) that the relation ${m_g}/{m_i}$ appears ``outside'' the gauge potential
$B^a{}_\mu$. This means essentially that, in this case, the gravitational field equations can
be consistently solved for $B^a{}_\mu$ independently of any test--particle property. This is
the fundamental reason for teleparallel gravity to remain as a viable theory for gravitation,
even in the absence of the weak equivalence principle. This result may have important
consequences for a fundamental problem of quantum gravity, namely, the conceptual difficulty
of reconciling the local character of the equivalence principle with the non--local character
of the uncertainty principle.\cite{chiao} Since teleparallel gravity can be formulated
independently of the equivalence principle, the quantization of the gravitational field may
possibly appear more consistent if considered in the teleparallel picture.

A further consequence that emerges from the conceptual differences between general relativity
and teleparallel gravity is that, whereas in the former curvature is used to geometrize the
gravitational interaction---spinless particles follow geodesics---in the latter torsion
describes the gravitational interaction by acting as a force---trajectories are not given by
geodesics, but by force equations. According to the teleparallel approach, therefore, the
role played by torsion is quite well defined: it appears as an alternative to curvature in
the description of the gravitational field, and is consequently related with the same degrees
of freedom of gravity. Now, this interpretation is completely different from that appearing
in more general theories, like Einstein--Cartan and gauge theories for the Poincar\'e and the
affine groups. In these theories, curvature and torsion are considered as independent fields,
related with different degrees of freedom of gravity, and consequently with different
physical phenomena. This is a conflicting situation as these two interpretations cannot be
both correct: if one is correct, the other is necessarily wrong.

We come then to the second part of the review, presented in Section~\ref{gauge}. In this
part, as an attempt to solve the above described paradox, we have critically reviewed the
physics of torsion in gravitation. More specifically, we have used the general covariance
principle---seen as an alternative version of the strong equivalence principle---to study the
gravitational coupling prescription in the presence of curvature and torsion. We recall that,
according to this principle, in order to make an equation generally covariant, a connection
is always necessary, which is in principle concerned only with the {\em inertial} properties
of the frame under consideration. Then, by using the equivalence between inertial and
gravitational effects, instead of representing inertial properties, this connection can
equivalently be assumed to represent a {\em true gravitational field}. In this way, equations
valid in the presence of gravitation can be obtained from the corresponding special relativity
equations. Now, as we have seen, the inertial compensating connection $\onehalf ( f_b{}^c{}_a
+ f_a{}^c{}_b - f^c{}_{ba})$, also known as {\em object of anholonomy}, is related to a
general Lorentz spin connection $A^c{}_{ab}$ through
\begin{equation}
\onehalf \left( f_b{}^c{}_a + f_a{}^c{}_b - f^c{}_{ba} \right) =
A^c{}_{ab} - K^c{}_{ab}.
\label{gecoprin}
\end{equation}
This equation is nothing but an expression of the equivalence principle: whereas its
left-hand side involves only {\it inertial} properties of a given frame, its right-hand side
contains purely {\it gravitational} quantities. Therefore, to replace inertial effects by a
true gravitational field means to replace the left-hand side by the right-hand side of
Eq.~(\ref{gecoprin}). This means essentially that the dynamical spin connection, that is, the
spin connection defining the covariant derivative, and consequently the gravitational
coupling prescription, is given by the right-hand side of Eq.~(\ref{gecoprin}). Even in the
presence of curvature and torsion, therefore, torsion appears as playing the role of
gravitational force. This result gives support to the point of view of teleparallel gravity,
according to which torsion does not represent additional degrees of freedom of gravity, but
simply an alternative way of representing the gravitational field. Furthermore, since
$A^c{}_{ab} - K^c{}_{ab} = \Abol^c{}_{ab}$, the ensuing coupling prescription will always be
equivalent with the coupling prescription of general relativity, a result that reinforces the
completeness of this theory.

It should be remarked that the object of anholonomy is sometimes replaced by the spin
connection $A^c{}_{ab}$ only. The resulting coupling prescription is the one usually assumed
to hold in Einstein--Cartan and other gauge theories for gravitation. Although it can be made
to satisfy the usual (passive) strong equivalence principle, this coupling prescription
clearly violates its active version. Furthermore, it presents some drawbacks, of which the
most important is, perhaps, the fact that, when used to describe the gravitational coupling
of the electromagnetic field, it violates the $U(1)$ gauge invariance of Maxwell's theory. On
the other hand, the coupling prescription implied by the general covariance principle
presents several formal advantages: it preserves the role played by torsion in teleparallel
gravity, it is consistent with both the active and passive versions of the strong equivalence
principle, it can be applied in the Lagrangian or in the field equations with the same
result, and when used to describe the interaction of the electromagnetic field with
gravitation, it does not violate the $U(1)$ gauge invariance of Maxwell's theory.

Summing up, we can say that the main output of the developments presented in this review is a
new interpretation for torsion in connection with the gravitational interaction. According to
this view, curvature and torsion are simply alternative ways of describing the gravitational
field. As a consequence, any gravitational phenomenon that can be interpreted in terms of
curvature, can also be interpreted in terms of torsion. Of course, we are aware that the
physical soundness of our arguments does not necessarily mean that they are correct, and that
a definitive answer can only be provided by experiments. However, considering that, at least
up to now, there are no compelling experimental evidences for {\it new physics associated
with torsion}, we could say that the teleparallel point of view is favored by the available
experimental data. For example, no new gravitational physics has ever been reported near a
neutron star. On the other hand, it is true that, due to the weakness of the gravitational
interaction, no experimental data exist on the coupling of the spin of the fundamental
particles to gravitation. Anyway, precision experiments,\cite{foot4} in
laboratory or as astrophysical and cosmological tests, are expected to be available in the
foreseeable future, when then a final answer will hopefully be achieved.

\begin{acknowledgments}
The authors would like to thank R. Aldrovandi, V. C. de Andrade, M. Cal\c cada, L. C. T.
Guillen, R. A. Mosna, T. Vargas, K. H. Vu, and C. M. Zhang for their participation in the
development of the ideas presented in this paper. They would like to thank also A. L. Barbosa and
Yu. N. Obukhov for useful discussions. This work was supported by FAPESP-Brazil, CNPq-Brazil,
CAPES-Brazil, and COLCIENCIAS-Colombia.
\end{acknowledgments}



\begin{thebibliography}{99}

\bibitem{exp}
The present experimental status of gravitation can be found in C. M. Will, {\it Living
Rev. Rel.} {\bf 4}, 4 (2001) [gr-qc/0103036].

\bibitem{mtw}
C. W. Misner, K. S. Thorne and J. A. Wheeler, {\it Gravitation} (Freeman, New York,
1973).

\bibitem{foot1}
The name teleparallel
gravity is normally used to designate the three-parameter teleparallel gravity. Here,
however, we use it as a synonymous of the teleparallel equivalent of general relativity, a
theory obtained for a specific choice of these parameters.

\bibitem{equiva}
See, for example, M. Schweizer and N. Straumann, {\it Phys. Lett.} {\bf A71}, 493
(1979); M. Schweizer, N. Straumann and A. Wipf, {\it Gen. Rel. Grav} {\bf 12}, 951
(1980); J. Nitsch and F. W. Hehl, {\it Phys. Lett.} {\bf B90}, 98 (1980).

\bibitem{paper1}
V. C. de Andrade and J. G. Pereira, {\it Phys. Rev.} {\bf D56}, 4689 (1997).

\bibitem{foot2}
Actually, a pseudo-Riemannian
structure, with signature +2. Despite not strictly correct, we will keep the word
``Riemannian'' to designate this kind of spacetime structure.

\bibitem{wep}
R. Aldrovandi, J. G. Pereira and K. H. Vu, {\it Gen. Rel. Grav.} {\bf 36}, 101 (2004).

\bibitem{synge}
For a critical discussion of the equivalence principle, see the preface of J. L. Synge,
{\it Relativity: The General Theory} (North-Holland, Amsterdam, 1960); see also T.
Damour, {\it Questioning the Equivalence Principle}, contribution to the ONERA workshop
on {\it Missions spatiales en physique fondamentale} (Chatillon, 18-19/1/2001), to
appear in a special issue of the {\it Comptes Rendus de l'Academie des Sciences
(Paris)}, ed. C. Bord\'e and P. Touboul [gr-qc/0109063].

\bibitem{quantu}
M. P. Haugan and C. L\"amerzahl, {\it Lect. Notes Phys.} {\bf 562}, 195 (2001);
C. L\"ammerzahl, {\it Gen. Rel. Grav.} {\bf 28}, 1043 (1996); C. L\"ammerzahl,
{\it Acta Phys. Pol.} {\bf 29}, 1057 (1998); C. L\"ammerzahl,
{\it Class. Quantum Grav.} {\bf 15}, 13 (1998).

\bibitem{hehl2}
F. H. Hehl, P. von der Heyde, G. D. Kerlick and J. M. Nester, {\it Rev. Mod.
Phys.} {\bf 48}, 393 (1976).

\bibitem{hehl2lec}
F. W. Hehl, J. Lemke and E. W. Mielke, {\em Two Lectures on Fermions and Gravity}, in
{\em Geometry and Theoretical Physics}, ed. by J. Debrus and A. C. Hirshfeld (Springer,
Heidelberg, 1991).

\bibitem{op1}
Yu. N. Obukhov and J. G. Pereira, {\it Phys. Rev.} {\bf D67}, 044016 (2003).

\bibitem{maluf3}
J. W. Maluf, {\it Phys. Rev.} {\bf D67}, 108501 (2003).

\bibitem{mielke}
E. W. Mielke, {\it Phys. Rev.} {\bf D69}, 128501 (2004).

\bibitem{op2}
Yu. N. Obukhov and J. G. Pereira, {\it Phys. Rev.} {\bf D69}, 128502 (2004).

\bibitem{tsc}
V. C. de Andrade, L. C. T. Guillen and J. G. Pereira, {\it Phys. Rev.} {\bf D64},
027502 (2001).

\bibitem{hammond}
R. T. Hammond, {\it Rep. Prog. Phys.} {\bf 65}, 599 (2002).

\bibitem{ecar}
See, for example, V. de Sabbata and M. Gasperini, {\it Introduction to Gravitation} (World
Scientific, Singapore, 1985).

\bibitem{kibble}
T. W. B. Kibble, {\it J. Math. Phys.} {\bf 2}, 212 (1961).

\bibitem{hcmn}
F. W. Hehl, J. D. McCrea, E. W. Mielke and Y. Ne'emann, {\it Phys. Rep.} {\bf 258}, 1
(1995).

\bibitem{mospe}
R. A. Mosna and J. G. Pereira, {\it Gen. Rel. Grav.} {\bf 36}, 2525 (2004) [gr-qc/0312093].

\bibitem{newt1}
H. I. Arcos and J. G. Pereira, {\it Class. Quant. Grav.} {\bf 21}, 5193 (2004)
[gr-qc/0408096].

\bibitem{newt2}
V. C. de Andrade, H. I. Arcos and J. G. Pereira, {\it Int. J. Mod. Phys.} {\bf D13}, 807
(2004).

\bibitem{koba}
S. Kobayashi and K. Nomizu, {\em Foundations of Differential Geometry}
(Interscience, New York, 1963).

\bibitem{livro}
R. Aldrovandi and J. G. Pereira, {\em An Introduction to Geometrical
Physics} (World Scientific, Singapore, 1995).

\bibitem{abp1}
R. Aldrovandi, P. B. Barros and J. G. Pereira, {\it Gen. Rel. Grav.} {\bf 35}, 991 (2003).

\bibitem{sauer1}
For a hystorical account, see T. Sauer, {\it Field equations in teleparallel spacetime:
Einstein's `Fernparallelismus' approach towards unified field theory} [physics/0405142];
see also L. O'Raifeartaigh and Straumann, {\it Rev. Mod. Phys.} {\bf 72}, 1 (2000) and
L. O'Raifeartaigh, {\em The Dawning of Gauge Theory} (Princeton University Press, Princeton,
1998).

\bibitem{utiyama}
R. Utiyama, {\it Phys. Rev.} {\bf 101}, 1597 (1956).

\bibitem{moller}
C. M{\o}ller, {\it K. Dan. Vidensk. Selsk. Mat. Fys. Skr.} {\bf 1} (No.10), 1 (1961).

\bibitem{pelle}
C. Pellegrini and J. Plebanski, {\it K. Dan. Vidensk. Selsk. Mat. Fys. Skr.} {\bf 2} (No.2), 1 (1962).

\bibitem{haya}
K. Hayashi and T. Nakano, {\it Prog. Theor. Phys.} {\bf 38}, 491 (1967).

\bibitem{hehl}
F. W. Hehl, in {\em Cosmology and Gravitation}, ed. by P. G.
Bergmann and  V. de Sabbata (Plenum, New York, 1980).

\bibitem{hehl0}
F. Gronwald and F. W. Hehl, {\it On the Gauge Aspects of Gravity}, in {\it Proceedings of
the 14th School of Cosmology and Gravitation}, Erice, Italy, ed. by P. G. Bergmann, V. de
Sabbata and H.-J Treder (World Scientific, Singapore, 1996)

\bibitem{blago}
M. Blagojevi\'c,
{\it Gravitation and Gauge Symmetries} (IOP Publishing, Bristol, 2002).

\bibitem{hayshi}
K. Hayashi and T. Shirafuji, {\it Phys. Rev.} {\bf D19}, 3524 (1979).

\bibitem{kopc}
W. Kopczy\'nski, {\it J. Phys.} {\bf A15}, 493 (1982).

\bibitem{azeredo}
R. de Azeredo Campos and C. G. Oliveira, {\it Nuovo Cimento} {\bf B74}, 83 (1983).

\bibitem{nitsch1}
F. M\"uller-Hoissen and J. Nitsch, {\it Gen. Rel. Grav.} {\bf 17}, 747 (1985).

\bibitem{mielke2}
E. W. Mielke, {\it Ann. Phys. (NY)} {\bf 219}, 78 (1992).

\bibitem{nes2}
R. S. Tung and J. M. Nester, {\it Phys. Rev.} {\bf D60}, 021501 (1999).

\bibitem{fi}
V. A. Fock and D. Ivanenko, {\it Z. Phys.} {\bf 54}, 798 (1929); V. A. Fock, {\it Z.
Phys.} {\bf 57}, 261 (1929).

\bibitem{dirac}
P. A. M. Dirac, in {\it Planck Festscrift}, ed. by W. Frank
(Deutscher Verlag der Wissen\-schaften, Berlin, 1958).

\bibitem{ramond}
See, for example, P. Ramond, {\it Field Theory: A Modern Primer}, 2nd edition
(Addison-Wesley, Redwood, 1989).

\bibitem{veltman}
M. J. G. Veltman, {\it Quantum Theory of Gravitation}, in {\it Methods in Field
Theory}, Les Houches 1975, ed. by R. Balian and J. Zinn-Justin (North-Holland, Amsterdam,
1976).

\bibitem{hamm1}
See, for example, R. T. Hammond, {\it Gen. Rel. Grav.} {\bf 28}, 749 (1996).

\bibitem{paper2}
V. C. de Andrade and J. G. Pereira, {\it Gen. Rel. Grav.} {\bf 30} 263 (1998).

\bibitem{sha}
See, for example, I. L. Shapiro, {\it Phys. Rep.} {\bf 357}, 113 (2002).

\bibitem{bdt}
I. M. Benn, T. Dereli and R. W. Tucker, {\it Phys. Lett.} {\bf B96}, 100 (1980).

\bibitem{sasi}
See, for example, V. de Sabbata and C. Sivaram, {\em Spin and Torsion in Gravitation}
(World Scientific, Singapore, 1994).

\bibitem{fhyo}
F. W. Hehl and Yu. N. Obukhov, in {\em Testing Relativistic Gravity in
Space: Gyroscopes, Clocks, Interferometers ...}, {\it Lect. Notes Phys.} {\bf 562}, 479
(2001).

\bibitem{vector}
V. C. de Andrade and J. G. Pereira, {\it Int. J. Mod. Phys.} {\bf D8}, 141 (1999).

\bibitem{mollersbook}
C. M{\o}ller, {\em The Theory of Relativity}, 2nd edition (Clarendon Press, Oxford, 1966).

\bibitem{maluf}
J. W. Maluf, {\it J. Math. Phys.} {\bf 35}, 335 (1994).

\bibitem{gemt}
V. C. de Andrade, L. C. T. Guillen and J. G. Pereira, {\it Phys. Rev. Lett.} {\bf 84},
4533 (2000).

\bibitem{belinfante}
F. J. Belinfante, {\it Physica} {\bf 6}, 687 (1939).

\bibitem{landau}
L. D. Landau and E. M. Lifshitz, {\it The Classical Theory of Fields}
(Pergamon, Oxford, 1975).

\bibitem{trautman}
See, for example, A. Trautman, in {\it Gravitation: an Introduction to Current
Research}, ed. by L. Witten (Wiley, New York, 1962).

\bibitem{others}
A. Papapetrou, {\it Proc. Roy. Irish Acad.} {\bf A52}, 11 (1948);
P. G. Bergmann and R. Thompson, {\it Phys. Rev.} {\bf 89}, 400 (1953);
C. M{\o}ller, {\it Ann. Phys. (NY)} {\bf 4}, 347 (1958).

\bibitem{recent}
M. Dubois-Violette and J. Madore, {\it Comm. Math. Phys.} {\bf 108}, 213 (1987);
L. B. Szabados, {\it Class. Quantum Grav.} {\bf 9}, 2521 (1992);
J. M. Aguirregabiria, A. Chamorro and K. S. Virbhadra, {\it Gen. Rel. Grav.} {\bf 28}, 1393
(1996); T. Shirafuji and G. L. Nashed, {\it Prog. Theor. Phys.} {\bf 98}, 1355 (1997);
S. Deser, J. S. Franklin and D. Seminara, {\it Class. Quantum Grav.} {\bf 16}, 2815 (1999);
S. V. Babak and L. P. Gri\-shchuck, {\it Phys. Rev.} {\bf D61}, 024038 (2000); Y. Itin
{\it Class. Quantum Grav.} {\bf 19}, 173 (2002); Y. Itin, {\it Gen. Rel. Grav.} {\bf 34}, 1819
(2002); J. W. Maluf, {\it Gen. Rel. Grav.} {\bf 30}, 413 (1998); J. W. Maluf, J. F. da
Rocha-Neto, T. M. L. Tor\'\i bio and K.H. Castello-Branco, {\it Phys. Rev.} {\bf D65}, 124001
(2002).

\bibitem{broyork}
J. D. Brown and J. W. York, {\it Phys. Rev.} {\bf D47}, 1407 (1993).

\bibitem{qlo}
C. C. Chang and J. M. Nester, {\it Phys. Rev. Lett.} {\bf 83}, 1897 (1999).

\bibitem{maluf95}
J. W. Maluf, {\it J. Math. Phys.} {\bf 36}, 4242 (1995).

\bibitem{shirafuji96}
T. Shirafuji, G. L. Nashed and Y. Kobayashi, {\it Prog. Theor. Phys.}
{\bf 96}, 933 (1996).

\bibitem{hay72}
K. Hayashi, Lett. Nuovo Cimento {\bf 5}, 529 (1972).

\bibitem{weinberg}
See, for example, S. Weinberg, {\em Gravitation and Cosmology}
(Wiley, New York, 1972).

\bibitem{kopov}
N. P. Konopleva and V. N. Popov, {\em Gauge Fields} (Harwood, New York, 1980).

\bibitem{weitz}
See, for example, R. Weitzenb\"ock, {\em Invariantentheorie} (Noordhoff, Gronningen, 1923),
page 356.

\bibitem{mg9}
V. C. de Andrade, L. C. T. Guillen and J. G. Pereira, {\it Teleparallel Gravity: An
Over\-view}, in {\it Proceedings of the Ninth Marcel Grossmann Meeting}, ed. by
V. G. Gur\-zadyan, R. T. Jantzen and R. Ruffini (World Scientific, Singapore, 2002)
[gr-qc/0011087].

\bibitem{nit80}
J. Nitsch and F. W. Hehl, {\it Phys. Lett.} {\bf B90} 98 (1980).

\bibitem{aud81}
J. Audretsch, {\it Phys. Rev.} {\bf D24} 1470 (1981).

\bibitem{cw}
I. Ciufolini and J. A. Wheeler, {\em Gravitation and Inertia}
(Princeton University Press, Princeton, 1995)

\bibitem{teof}
J. G. Pereira, T. Vargas and C. M. Zhang, {\it Class. Quantum Grav.} {\bf 18} 833
(2001).

\bibitem{global}
R. Aldrovandi, J. G. Pereira and K. H. Vu, {\it Class. Quantum Grav.} {\bf 21}, 51
(2004).

\bibitem{wy}
T. T. Wu and C. N. Yang, {\it Phys. Rev.} {\bf D12}, 3845 (1975).

\bibitem{cow}
R. Colella, A. W. Overhauser and S. A. Werner, {\it Phys. Rev. Lett.} {\bf 34}, 1472
(1974).

\bibitem{gabere}
A. K. Lawrence, D. Leiter and G. Samozi, {\it Nuovo Cimento}
{\bf 17B}, 113 (1973); L. H. Ford and A. Vilenkin, {\it J. Phys.} {\bf A14}, 2353 (1981);
V. B. Bezerra, {\it Class. Quantum Grav.} {\bf 8}, 1939 (1991).

\bibitem{foot3}
We remark that absence of torsion, which happens in internal gauge theories, is different
from the presence of a vanishing torsion, which happens in general
relativity.

\bibitem{sciama}
D. W. Sciama, {\it Rev. Mod. Phys.} {\bf 36} 463 (1964).

\bibitem{lorentz}
M. Cal\c cada and J. G. Pereira, {\it Int. J. Theor. Phys.} {\bf 41} 729 (2002).

\bibitem{singer}
I. M. Singer, {\it Comm. Math. Phys.} {\bf 60}, 7 (1978).

\bibitem{drech80}
S. K. Wong, {\it Nuovo Cimento} {\bf A65}, 689 (1970);
W. Drechsler, {\it Phys. Lett.} {\bf B90}, 258 (1980).

\bibitem{by}
K. Yee and M. Bander, {\it Phys. Rev.} {\bf D48}, 2797 (1993).

\bibitem{papapetrou}
A. Papapetrou, {\it Proc. R. Soc. (London)} {\bf A209}, 248 (1951).

\bibitem{chiao}
R. Y. Chiao, in {\it Wheeler's 90th Birthday Symposium Proceedings} (Cambridge University
Press, Cambridge, 2003) [gr-qc/0303100].

\bibitem{foot4}
A discussion, with a list of references on possible experiments to probe torsion, can be found
in Ref.~[15].

\end{thebibliography}
\end{document}